\documentclass[12pt,preprint]{aastex}

\newcommand{\ms}{\mbox{m s$^{-1}~$}}

\newcommand{\logrhk}{$\log R'_{\textrm{\tiny{HK}}}$~}
\newcommand{\logrhkm}{\log R'_{\textrm{\tiny{HK}}}}
\newcommand{\smw}{$S_{\textrm{\tiny MW}}~$}
\newcommand{\smi}{$S_{\tiny{\textrm{MIKE}}}~$}

\lefthead{Arriagada {\it et~al.\/}}
\slugcomment{Draft to be submitted to ApJ}
\begin{document}

\title{ Chromospheric activity of Southern Stars from the Magellan Planet Search Program\altaffilmark{1}}

\author{Pamela Arriagada\altaffilmark{2},
}

\authoremail{parriaga@astro.puc.cl}

\altaffiltext{1}{Based on observations obtained with
the Magellan Telescopes, operated by the Carnegie
Institution, Harvard University, University of Michigan,
University of Arizona, and the Massachusetts Institute
of Technology.} 

\altaffiltext{2}{Department of Astronomy, Pontificia
Universidad Cat\'olica de Chile, Casilla 306, Santiago 22, Chile}


\begin{abstract}
I present chromospheric activity measurements of $\sim$ 670 F, G, K and M main sequence stars in the Southern Hemisphere, from $\sim$8000 archival high-resolution echelle spectra taken at Las Campanas Observatory since 2004. These stars were targets from the Old Magellan Planet Search, and are now potential targets for the New Magellan Planet Search that will look for rocky and habitable planets. Activity indexes ($S$-values) are derived from Ca II H \& K line cores and then converted to the Mt. Wilson system. From these measurements, chromospheric  (\logrhk) indexes are derived, which are then used as indicators of the level of radial-velocity jitter, age and rotation periods these stars present.
\end{abstract}

\keywords{stars:activity, stars:chromospheres, stars:fundamental parameters, stars:rotation} 

\section{Introduction}
\label{intro}
The Magellan Planet Search Program has been monitoring radial velocities of 690 G, F and K main sequence stars since December 2002 at high-spectral resolution with the main goal of finding extra solar planets. 
Precision Doppler surveys rely on finding a repeating pattern on the radial velocities, and thus, one must take into account major sources of error that may mimic a planet's signature. One of these sources is the activity in the star's chromosphere, otherwise called ``photospheric jitter'' (Queloz et al. 2001; Henry, Donahue, \& Baliunas 2002; Santos et al. 2003). Activity measurements are, therefore, essential for selecting the most inactive, stable stars for a planet search survey.
Magnetic activity levels can be determined from the strength of emission in the Ca II H \& K line cores of a star's spectrum, which gives an estimate of the stellar jitter and the rotation period, both critical values for understanding and interpreting the noise present in radial velocity measurements (Noyes et al. 1984; Saar \& Fischer 2000; Santos et al. 2000).
Since 1966, the Mount Wilson program has been monitoring Ca II H \& K emission of more than 1200 Northern-Hemisphere dwarfs and giants, defining the Mount Wilson S-value (\smw) (Duncan et al. 1991). Due to the long term monitoring of their stars, the \smw has been used by other programs as the standard metric of photospheric activity.
In the Southern-Hemisphere, two large programs have made the efforts to provide measurements and analysis of photospheric activity of solar-type stars. Henry et al. (1996) observed more than 800 stars at CTIO using the Cassegrain Spectrograph at the 1.5m telescope. More recently, Gray et al (2006) has observed more than 1600 dwarf and giants as part of the Nstars project. Although they provide us with a useful testbed to the conclusions drawn in the north, and also an independent sample for statistical analysis, both were done using low-resolution spectra and none provide the long-term coverage as Mount Wilson's. Other efforts in the Southern Hemisphere were made by Tinney et al. (2002) and Jenkins et al. (2006), who performed a long-term analysis of $\sim$200 stars as part of the Anglo-Australian Planet Search (AAPS) using high-resolution spectra, as well as chromospheric measurements of 353 bright stars by Jenkins et al. (2008) . 

\section{Observations and data reduction}

The Magellan Planet Search Program has been monitoring $\sim$690 stars since late 2002. The observations have been done using the MIKE echelle spectrograph (Bernstein et al. 2003) mounted on the 6.5-m Clay Telescope (Magellan II) at Las Campanas Observatory. Using a 0.35 arc-sec slit, MIKE obtains spectra with a resolution of R $\sim$ 70000 in the blue and $\sim$50000 in the red, covering the wavelength range from 3900--6200 \AA~divided into a red and a blue CCD. The Iodine spectrum (5000 - 6200 \AA) falls on the red CCD while the blue CCD captures the Ca II H and K lines to monitor stellar activity. The detector on MIKE is a Lincoln Labs 2048$\times$4096 15$\mu$m CCD , with a Ca II HK slit-to-detector efficiency of $\sim$30 per cent at 3800 \AA. 

The targets presented here are part of the Magellan Planet Search Program stars, which are 690 F7-M5 dwarfs and subgiants, that were observed in seeing that ranged from 0.5 to 1.5 arcsec. The dispersion at the Ca II H \& K lines is 0.02 \AA ~pixel$^{-1}$. Exposure times ranged from 90 seconds on brightest objects to 600 on the fainter ones, giving a SNR per pixel that would fall between 20 and 100 at the Ca II H \& K lines.

Extraction from raw CCD images was carried out using a modified version of the pipeline used to extract spectra from the red CCD images used then to calculate velocities. Besides the standard reduction, this pipeline also measures scattered light from inter order pixels and subtracts it. No sky subtraction is done as the sky brightness around the Ca II H \& K lines (and throughout the whole echellogram) is negligible compared to the brightness of our sources. A cosmic ray removal is also performed on the two dimensional echellogram, as with all MIKE data. Wavelength calibration was done using ThAr spectra acquired each night to focus the instrument. 

Since the blaze function removal is not part of the standard pipeline, and no flux standards are observed during the Planet Search program, a different approach had to be applied. In order to correct for the blaze function we fit a 7th order polynomial to a quartz lamp spectrum, and divided it out of all spectra in the region of interest. I note at this point that during the first nine observing runs quartz lamp spectra were not properly taken and a slightly modified reduction code was used in order to extract the spectra. 

Wavelength calibration was done using ThAr spectra that was acquired each night to focus the instrument and to set the grating in place. All stars were then cross-correlated with a binned spectrum of the sun taken from the NSO, to correct for the the barycentric velocity of each star.


\section{Analysis}

\subsection{Derivation of S indices from MIKE/Magellan observations}

Duncan et al. (1991) defines the $S$-index as the ratio between the total counts  $H$ and $K$ bandpasses centered on the H and K lines, and the total counts in the $R$ and $V$ bandpasses centered in the continuum. The MW project calculated this index individually each night using the HKP-2 spectrometer, which is a specialized multichannel spectrometer. The $H$ and $K$ bandpasses, have triangular profiles with full width at half maximums of 1.09 \AA~and are centered at the very cores of the H and K lines (3933.667 and 3968.470 \AA~respectively). The other two, $R$ and $V$, have rectangular profiles with widths of 20 \AA~and are centered in the continuum at 3901 and 4001 \AA.

\begin{equation}
S_{\tiny{\textrm{MW}}}=\alpha\frac{N_{H}+N_{K}}{N_{R}+N_{V}}
\end{equation}

\noindent  $\alpha$ is a constant that was calculated to be 2.4, and that would make the mean $S$ correspond to the mean $F$ of standard stars, the original chromospheric measurements determined at MW using the old HKP-1 spectrometer which had different bandpasses.

Following the prescription of Duncan et al (1991), I have simulated the measurement of the Mount Wilson spectrometers by defining two triangular bandpasses with 1.09 \AA~FWHM centered on the H and K lines, and two rectangular channels 20 \AA~wide centered at  3901 and 4001 \AA. Figure 1 shows the position of these channels in our MIKE spectra.
I have summed counts within these four effective bandpasses and taken the ratio between the sum of H and K and the sum of R and V. The relation used to determine MIKE $S$ values is :

\begin{equation}\label{lala}
S_{\tiny{\textrm{MIKE}}}=\frac{N_{H}+N_{K}}{N_{R}+N_{V}}
\end{equation}

In order to correct for systematic effects due in part to differences in removing the blaze function as noted in Section 2, and to upgrades of the MIKE blue CCD chip, which presented linearity problems between May 2004 and September 2005, we performed linear calibrations between the two earlier epochs and the latter one, shown in Figure 2, in the same way that MIKE S values are then calibrated into the Mount Wilson system as detailed in the next section.


\noindent Thus, using equation \ref{lala} we produced a set of \smi~that were then calibrated into the Mount Wilson System. Individual $S$ values will be available online. 

\subsection{Converting from \smi~to Mount Wilson \smw}

A conversion had to be applied in order to calibrate MIKE $S$ values into the Mount Wilson standard frame. Tinney et al. (2002), Jenkins et al. (2006) and Jenkins et al. (2008) have demonstrated that to bring $S$ values derived from high resolution spectra using the Duncan et al. (1991) prescription into the MW system only requires a linear calibration. To find this relation, a statistically significant amount of stars with \smw values must be used. Only two stars previously observed at MW were observed during the program (HD 10700 and HD 119217), so a different approach had to be taken in order to perform the calibration. 


Stars which have already been part of low and high resolution chromospheric studies and that were calibrated into the MW system were selected. 195 stars had already been observed by Gray et al. (2006)(G06), 117 by Henry (1996) (H96), 115 by Jenkins et al. (2008)(J08) and 7 by Wright et al. (2004)(W04). Figure 3 shows the median \smi value against the reported values from G06, H96, J08 and W04. A tight correlation is seen between \smi and all four studies, although the scatter increases when comparing against values obtained from low resolution spectra (G06, H96) and where the number of points is low (W04). Based on this result, we decided to use only the stars from J08, giving a linear least squares fit that has a slope of 1.049$\pm$0.05 and a zero point offset of 0.012$\pm$0.004. 
The derived calibrated S values are presented in Table 1, including both stars from Duncan et al. (1991). The values derived in this study agree within uncertainties.

\subsection{Uncertainties and random errors}

Since chromospherically-based quantities depend on intrinsic variability, which fluctuates on all timescales such as that of activity cycles or rotational periods, our final \logrhk estimations represent median values of the points we obtained during the period they were observed, but not true averages for the stars.

Measurement errors are caused by quality of the spectra as well as the quality of the calibration into the MW values. The 0.017 scatter shown in Figure 3 between the J08 data set is in part due to the stellar variability, since MIKE data are not contemporaneous with the
J08 data and many stars are in different parts of
their activity cycles. Measurements
for stars observed more frequently and for the full duration of
the Planet Search program will have correspondingly lower
uncertainties.

Previous works (Wright et al. 2004, Jenkins et al. 2006) have used measurements of the stable star $\tau$ Ceti (HD 10700) as a proxy for the random errors introduced
in the reduction procedure (scattered light removal, blaze function correction, cosmic ray removal and barycentric velocity correction). Considering that, first, $\tau$ Ceti has been shown to
be extremely stable (Baliunas et al. 1995a), and second, because it is an excellent source with which to search for any systematic errors in our precision velocities, this star has been observed continuously throughout the program; $\tau$ Ceti
serves as an excellent diagnostic star. 127 useful observations of $\tau$ Ceti have been acquired, as shown in Figure 5, with a standard deviation of 4.9\%, which is similar to that of the Keck and Lick errors from Wright et al. (2004).  





\subsection{\logrhk}

The $S$ index includes both photospheric and chromospheric information. However, for an activity analysis, we are only interested in the chromospheric component. The photospheric contribution, which depends on the stellar temperature (measured as $(B-V)$), is removed following the methodology by Noyes et al. (1984) in order to generate the \logrhk which appears in Table 1.  This transformation has been well calibrated for $0.44<(B-V)<0.9$, which means that for stars redder than 0.9, the calibration has greater uncertainties. 

Figure 4 shows \logrhk values from this study plotted against values from J08. The line denotes a 1:1 relationship, and visual inspection reveals a good match between the two sets. Since calibration stars were only observed once or twice by J08, we expected a high dispersion in the data due to propagated errors, such as the intrinsic variability of the stars, scatter in J08 values and scatter in my own values, which were discussed in the previous section.
In order to test if there is good agreement between the samples, first, I compare my values with J08 using

\begin{equation}\label{sigma}
\sigma=\frac{\log R'_{\textrm{\tiny HK,MIKE}}-\log R'_{\textrm{\tiny HK,FEROS}}}{\log R'_{\textrm{\tiny HK, MIKE}}}
\end{equation}

\noindent to look for any systematic differences. These values are plotted in Figure 4, where we see no clear trend. The mean value for $\sigma$ is -0.00013.  A two-sample Kolmogorov-Smirnov test was also performed to check for any difference between both data set distributions, obtaining a statistical estimator $D$=0.124 and a corresponding probability estimator of $P=$0.3, meaning that my sample is not significantly different from J08.

\subsection{Rotation periods and Ages}

Using the results from the activity analysis, one can also compute rotation periods and ages using empirically derived \logrhk relationships. 
From Mamajek \& Hillebrand (2008)



\begin{equation}\label{rotation}
\textrm{Ro} = \left\{ \begin{array}{lr}
 (0.233\pm0.015)-(0.689\pm0.063)(\logrhkm + 4.23) &\mbox{ if  \logrhk$ > -4.3$} \\
 (0.808\pm0.014)-(2.966\pm0.098)(\logrhkm + 4.52) &\mbox{ if  $-5.0 < $ \logrhk $< -4.3$}
       \end{array} \right.
\end{equation}


\noindent where Ro=$\log(P_{\textrm{rot}}/\tau)$ and $\tau$ is the convective turnover time as adopted from Noyes et al. (1984),

\begin{equation}\label{tau}
\log \tau = \left\{ \begin{array}{lr}
 1.362-0.166x+0.025x^2-5.323x^3 &\mbox{ if $x>0$} \\
 1.362-0.14x  &\mbox{ if $x<0$}
       \end{array} \right.
\end{equation}

\noindent where $x=1-(B-V)$ and the ratio of mixing length to scale height is 1.9. Noyes et al. (1984) note that this relationship has a rms about the mean curve of $\sim0.08$, which means that predicted rotational periods using equations \ref{rotation} and \ref{tau} may be fairly accurate.



To estimate ages for individual stars, we use the age-chromospheric activity relation of Mamajek \& Hillebrand (2008):

\begin{equation}\label{age}
\log t=-38.053 -17.912\logrhkm-1.6675(\logrhkm)^2
\end{equation}

\noindent where $t$ is the stellar age in years. Mamajek \& Hillebrand (2008) report an rms of $\sim$0.07 dex in $\log t/$yr for stars with -5.1$<$ \logrhk $<$ -4.3. I have extrapolated this relation to lower \logrhk in order to get an estimate of the ages, therefore for less active stars the estimated age will not be as accurate. I should also note at this point that $R'_{\textrm{\tiny HK}}$ corresponds to the activity level averaged over many activity cycles. For this reason, one should obtain as many observations as possible in order to derive a value closer to the actual age of the star using this relation, which is not always the case for the program stars. 




I present a histogram of the distribution of the median activity levels for the Magellan target stars in Figure 6.  The bimodal distribution of activity in \logrhk noted previously by Gray et al. (2003), Jenkins et al. (2006, 2008) is evident, although less pronounced. This slight discrepancy might be explained by the dependency of bimodality with metallicity ([M/H]), as shown by Gray et al. (2006). Another work, by Hall et al. (2007), shows a similar distribution to our sample.

\subsection{Radial velocity jitter}

Since intrinsic stellar variations can lead to false planet detections, when it shows periodic signals, or to non detections, when jitter is larger than planetary signature; it is essential to quantify this source in order to take it into account when searching for planets. Saar, Butler \& Marcy (1998), Santos et al. (2000) derived the first empirical relations detected RV dispersion, or jitter ($\sigma'_{\tiny{RV}}$), and the activity indices \logrhk for F,G and K dwarfs with only a small sample of stars. 

Wright (2005) 
used $\sim  450$ stars from the California and Carnegie Planet Search to derive empirical models to predict a star's radial velocity jitter based on its $(B-V)$ color, activity level and absolute magnitude. Since \logrhk is not well calibrated in the $B-V<0.4$ or $B-V>0.9$, an alternative metric of stellar activity, $F_{\tiny(Ca II)}$ is used instead. 
Updated relations to estimate the baseline expected jitter as a function of excess activity $\Delta S$ and $B-V$ colors were recently published by Isaacson \& Fischer (2010) using more than 2600 stars from the California Planet Search. 

Using \emph{Hipparcos} (Perryman et al. 1997) colors and parallaxes, I have applied both Wright's empirical model and Isaacson's relations to my calculated activity S-values to predict the expected radial velocity jitter of the Magellan program stars. These results are given in the last two columns in Table 1.

\section{Conclusions}


Chromospheric activity and $S$-values from over 9,000
archival spectra from the Magellan Planet Search Program, taken over a baseline of 6 years, have been measured.  The spectra were taken using the MIKE spectrograph at Las Campanas Observatory. 


Analysis of the measured level of activity of $\tau$ Ceti yields a random error of $\sim$6\% and the S-values presented here correspond to the median activity levels of each star during the time span of the observations.

As stellar activity has proven to mimic an extrasolar planet signal, an expected value of the RV jitter due to the star's intrinsic variability is an essential tool when it comes to selecting targets, as it can constrain the minimum RV amplitude variation detectable due to a true planetary companion. I have computed the expected stellar RV jitter for all the stars in the sample, with the main goal of selecting the best targets for the new Magellan Planet Search, which is making use of the new Planet Finder Spectrograph (Crane et al. 2006; Crane et al. 2008).  Advantages of this new spectrograph include
higher throughput, higher resolution, active and passive temperature
stabilization, fixed format, and all optics optimized for the
Iodine region (5000 to 6200 Angstroms).  Data collected in the first five months of scientific operation indicate that velocity precision better than 1\ms RMS is being achieved (Crane et al. 2010).




\acknowledgements
I am grateful to the Magellan observers : R.P. Butler, Dante Minniti and Merecedes L\'opez-Morales. I thank James S. Jenkins for the helpful discussions. I also acknowledge support by the FONDAP Center for Astrophysics
15010003, BASAL CATA Center for Astrophysics and Associated
Technologies PFB-06, and MIDEPLAN Iniciativa Cient\'ifica Milenio
project Milky Way Millenniun. 
This paper has made use of the Simbad and NASA ADS data bases.



\clearpage

\begin{deluxetable}{lcccccccc}
\tablecaption{Derived S-values from MIKE spectra converted to MW system,  $S_{\textrm {\tiny MIKE}}$; chromospheric activity indices, \logrhk; rotation periods, P$_{\textrm {\tiny rot}}$; ages, $\log$(Age/years); and estimated jitter, $\sigma'_{\textrm {\tiny RV}}$ (Isaacson \& Fischer (2010) and Wright (2005)) labeled as 1 and 2 respectively.}
\label{stellparams}
\tablewidth{0pt}
\tablehead{
Name & $B-V$ & $N_{\textrm {\tiny obs}}$ & $S_{\textrm {\tiny MIKE}}$ & \logrhk & P$_{\textrm {\tiny rot}}$ & $\log$(Age/years) & $\sigma'_{\textrm {\tiny RV 1}}$& $\sigma'_{\textrm {\tiny RV 2}}$\\
(HD) &&&&&(days)&&(\ms)&(\ms)
}
\startdata
224789  &   0.863  &       7&0.458 & -4.47 &14. &8.71 &3.5 &8.8  \\ 
225155  &   0.741  &       4&0.150 & -5.08 &... &9.91 &2.1 &2.2  \\ 
225299  &   0.710  &       6&0.193 & -4.86 &28. &9.61 &2.3 &2.1  \\ 
23  &   0.577  &       8&0.173 & -4.89 &15. &9.67 &2.8 &3.6  \\ 
55  &   1.076  &       4&0.454 & ... &... &... &1.6 &2.1  \\ 
361  &   0.624  &       6&0.195 & -4.81 &17. &9.52 &3.1 &4.6  \\ 
798  &   0.448  &       6&0.157 & -4.97 &5. &9.78 &2.4 &3.0  \\ 
1002  &   0.640  &       4&0.143 & -5.12 &... &9.92 &2.2 &2.4  \\ 
1237  &   0.750  &       1&0.458 & -4.36 &6. &8.33 &3.5 &18.4  \\ 
hip1532  &   1.318  &       3&0.839 & ... &... &... &2.4 &2.1  \\ 
1690  &   1.354  &       2&0.166 & ... &... &... &0.5 &2.2  \\ 
1893  &   0.769  &      11&0.244 & -4.73 &26. &9.36 &2.5 &4.4  \\ 
1910  &   1.075  &       6&0.952 & ... &... &... &1.6 &7.1  \\ 
2222  &   0.675  &       7&0.139 & -5.16 &... &9.97 &2.1 &2.2  \\ 
3222  &   0.854  &       7&0.189 & -4.95 &44. &9.75 &2.2 &2.1  \\ 
3359  &   0.780  &       5&0.144 & -5.12 &... &9.93 &2.1 &2.1  \\ 
4113  &   0.716  &       9&0.149 & -5.08 &... &9.91 &2.1 &2.1  \\ 
4333  &   0.414  &       4&0.175 & -4.86 &3. &9.62 &2.4 &4.0  \\ 
4631  &   0.774  &       7&0.159 & -5.04 &... &9.86 &2.1 &2.1  \\ 
5499  &   0.987  &       3&0.114 & ... &... &... &1.6 &4.3  \\ 
5349  &   0.991  &       2&0.115 & ... &... &... &1.6 &4.3  \\ 
6107  &   0.650  &       8&0.143 & -5.12 &... &9.93 &2.2 &2.4  \\ 
hip4845  &   1.365  &       3&1.270 & ... &... &... &3.5 &2.2  \\ 
6156  &   0.800  &       5&0.355 & -4.54 &17. &8.90 &3.0 &8.0  \\ 
6236  &   0.591  &       4&0.146 & -5.08 &... &9.91 &2.3 &2.6  \\ 
6434  &   0.613  &       8&0.144 & -5.10 &... &9.90 &2.2 &2.5  \\ 
6790  &   0.561  &       5&0.154 & -5.00 &... &9.82 &2.5 &2.7  \\ 
6880  &   0.764  &       4&0.234 & -4.75 &27. &9.41 &2.5 &4.1  \\ 
6910  &   0.651  &       6&0.126 & -5.29 &... &10.14 &1.9 &4.3  \\ 
7134  &   0.589  &       6&0.151 & -5.04 &... &9.87 &2.4 &2.6  \\ 
7399  &   0.476  &       8&0.134 & -5.17 &... &9.99 &2.1 &3.0  \\ 
7449  &   0.575  &       8&0.168 & -4.92 &15. &9.71 &2.7 &2.6  \\ 
7661  &   0.753  &       4&0.397 & -4.43 &10. &8.59 &3.3 &13.4  \\ 
7786  &   0.494  &       6&0.171 & -4.88 &7. &9.64 &2.8 &5.0  \\ 
8049  &   0.876  &       8&0.678 & -4.30 &3. &8.14 &4.5 &16.2  \\ 
8076  &   0.622  &      14&0.220 & -4.71 &14. &9.32 &3.6 &6.9  \\ 
8129  &   0.702  &       4&0.180 & -4.91 &29. &9.70 &2.2 &2.1  \\ 
8326  &   0.970  &       7&0.340 & ... &... &... &2.7 &2.1  \\ 
8406  &   0.656  &       4&0.162 & -4.98 &27. &9.80 &2.5 &2.3  \\ 
8581  &   0.569  &       7&0.133 & -5.20 &... &10.04 &2.1 &2.7  \\ 
9175  &   0.656  &       7&0.145 & -5.10 &... &9.90 &2.2 &2.3  \\ 
9782  &   0.593  &       7&0.141 & -5.12 &... &9.93 &2.2 &2.6  \\ 
9847  &   0.712  &       5&0.140 & -5.15 &... &9.97 &2.0 &2.5  \\ 
9905  &   0.761  &       8&0.172 & -4.97 &38. &9.78 &2.2 &2.1  \\ 
hip7554  &   1.416  &       2&1.782 & ... &... &... &4.8 &2.3  \\ 
10008  &   0.797  &       3&0.421 & -4.45 &11. &8.62 &3.4 &11.3  \\ 
10226  &   0.607  &       3&0.176 & -4.89 &18. &9.66 &2.8 &2.5  \\ 
10576  &   0.590  &       9&0.142 & -5.11 &... &9.91 &2.2 &2.6  \\ 
10370  &   0.714  &       5&0.164 & -5.00 &35. &9.81 &2.2 &2.1  \\ 
10519  &   0.617  &       4&0.138 & -5.16 &... &9.98 &2.1 &2.5  \\ 
10678  &   0.714  &       7&0.213 & -4.79 &25. &9.48 &2.4 &4.1  \\ 
10611  &   0.853  &       7&0.477 & -4.44 &12. &8.61 &3.6 &10.0  \\ 
10700  &   0.727  &     127&0.160 & -5.02 &... &9.84 &2.1 &2.1  \\ 
11264  &   0.664  &       3&0.140 & -5.15 &... &9.97 &2.1 &2.2  \\ 
11231  &   0.463  &       7&0.127 & -5.26 &... &10.10 &1.9 &3.0  \\ 
11754  &   0.682  &       6&0.137 & -5.17 &... &9.99 &2.1 &2.6  \\ 
12058  &   1.143  &       4&0.944 & ... &... &... &1.6 &4.5  \\ 
13060  &   0.797  &       5&0.217 & -4.82 &33. &9.55 &2.4 &2.1  \\ 
12951  &   0.576  &       9&0.130 & -5.23 &... &10.08 &2.0 &2.6  \\ 
12585  &   0.662  &      11&0.157 & -5.02 &... &9.85 &2.4 &2.3  \\ 
12617  &   1.020  &       5&0.510 & ... &... &... &1.6 &4.1  \\ 
13350  &   0.766  &       4&0.137 & -5.17 &... &9.99 &2.0 &2.6  \\ 
13386  &   0.880  &       5&0.174 & -5.02 &... &9.84 &2.1 &2.1  \\ 
13724  &   0.667  &       8&0.205 & -4.79 &21. &9.48 &3.3 &4.5  \\ 
13808  &   0.870  &       4&0.235 & -4.83 &37. &9.56 &2.4 &2.1  \\ 
hip10337  &   1.346  &       2&1.689 & ... &... &... &4.6 &2.2  \\ 
13789  &   1.055  &       4&0.711 & ... &... &... &1.6 &5.3  \\ 
14629  &   1.031  &       3&0.569 & ... &... &... &1.6 &4.5  \\ 
14635  &   1.079  &       4&0.671 & ... &... &... &1.6 &4.2  \\ 
14758  &   0.644  &       5&0.158 & -5.01 &... &9.83 &2.5 &2.3  \\ 
15234  &   0.417  &       7&0.170 & -4.89 &3. &9.66 &2.4 &3.0  \\ 
15507  &   0.670  &       5&0.161 & -5.00 &29. &9.82 &2.5 &2.2  \\ 
15590  &   0.652  &       4&0.131 & -5.23 &... &10.07 &2.0 &2.6  \\ 
15767  &   0.935  &       3&0.444 & ... &... &... &3.3 &5.4  \\ 
16348  &   1.045  &       5&0.524 & ... &... &... &1.6 &3.7  \\ 
16950  &   0.616  &       6&0.128 & -5.26 &... &10.11 &2.0 &2.6  \\ 
17155  &   1.040  &       3&0.294 & ... &... &... &1.6 &2.1  \\ 
17289  &   0.588  &       3&0.139 & -5.14 &... &9.96 &2.2 &2.6  \\ 
17134  &   0.634  &       5&0.143 & -5.12 &... &9.92 &2.2 &2.4  \\ 
hip12961  &   1.389  &       2&1.336 & ... &... &... &3.6 &2.3  \\ 
hip13389  &   1.549  &       1&0.255 & ... &... &... &1.3 &2.5  \\ 
18168  &   0.933  &       3&0.516 & ... &... &... &3.7 &7.0  \\ 
18754  &   0.736  &       6&0.134 & -5.19 &... &10.02 &2.0 &4.3  \\ 
18708  &   0.651  &       5&0.134 & -5.20 &... &10.03 &2.0 &2.3  \\ 
18819  &   0.597  &       5&0.120 & -5.37 &... &10.23 &1.8 &3.0  \\ 
18894  &   0.598  &       5&0.179 & -4.87 &16. &9.62 &2.9 &3.9  \\ 
19493  &   0.695  &       2&0.135 & -5.19 &... &10.02 &2.0 &2.6  \\ 
19330  &   0.566  &       8&0.181 & -4.84 &13. &9.59 &2.9 &4.7  \\ 
19880  &   0.715  &       4&0.134 & -5.20 &... &10.03 &2.0 &2.6  \\ 
20003  &   0.771  &       6&0.187 & -4.91 &36. &9.70 &2.3 &2.1  \\ 
19603  &   0.603  &       6&0.142 & -5.12 &... &9.92 &2.2 &2.5  \\ 
hip14593  &   1.340  &       2&0.778 & ... &... &... &2.2 &2.2  \\ 
19916  &   0.585  &       7&0.137 & -5.15 &... &9.97 &2.2 &2.6  \\ 
20155  &   0.622  &       6&0.139 & -5.15 &... &9.97 &2.1 &2.6  \\ 
20299  &   0.428  &       4&0.161 & -4.94 &4. &9.74 &2.3 &3.0  \\ 
20407  &   0.586  &      12&0.151 & -5.04 &... &9.87 &2.4 &2.6  \\ 
20339  &   0.614  &       6&0.143 & -5.11 &... &9.91 &2.2 &2.5  \\ 
20584  &   0.584  &       9&0.123 & -5.33 &... &10.19 &1.9 &3.8  \\ 
20657  &   0.800  &       5&0.136 & -5.18 &... &10.00 &2.0 &3.8  \\ 
21036  &   0.730  &       4&0.290 & -4.60 &17. &9.05 &2.8 &7.8  \\ 
21411  &   0.716  &       7&0.232 & -4.73 &22. &9.36 &2.5 &5.0  \\ 
21749  &   1.130  &       6&0.589 & ... &... &... &1.6 &2.1  \\ 
21841  &   0.666  &       8&0.235 & -4.69 &17. &9.27 &3.8 &6.6  \\ 
21938  &   0.591  &       6&0.143 & -5.11 &... &9.91 &2.2 &2.6  \\ 
22177  &   0.717  &       6&0.150 & -5.08 &... &9.90 &2.1 &2.1  \\ 
22323  &   0.401  &       5&0.222 & -4.66 &2. &9.21 &3.1 &16.0  \\ 
22582  &   0.730  &       5&0.148 & -5.09 &... &9.92 &2.1 &2.1  \\ 
23295  &   1.089  &       3&0.376 & ... &... &... &1.6 &2.1  \\ 
23576  &   0.587  &       6&0.141 & -5.12 &... &9.92 &2.2 &2.6  \\ 
24085  &   0.595  &       6&0.148 & -5.07 &... &9.89 &2.3 &2.6  \\ 
24331  &   0.913  &       7&0.273 & ... &... &... &2.6 &2.1  \\ 
24650  &   0.569  &       6&0.140 & -5.13 &... &9.94 &2.2 &2.7  \\ 
hip18280  &   1.366  &       5&0.992 & ... &... &... &2.7 &2.2  \\ 
25054  &   0.536  &       5&0.154 & -5.00 &... &9.82 &2.5 &2.7  \\ 
25004  &   1.214  &       5&1.221 & ... &... &... &1.6 &4.0  \\ 
25797  &   0.422  &       6&0.207 & -4.71 &2. &9.32 &3.1 &11.9  \\ 
26071  &   0.742  &      16&0.152 & -5.07 &... &9.90 &2.1 &2.6  \\ 
26040  &   0.576  &       4&0.152 & -5.03 &... &9.85 &2.4 &2.6  \\ 
hip19976  &   1.602  &       3&0.867 & ... &... &... &2.4 &2.5  \\ 
27426  &   1.052  &      13&0.671 & ... &... &... &1.6 &5.0  \\ 
hip20142  &   1.654  &       2&1.040 & ... &... &... &1.6 &2.5  \\ 
27466  &   0.673  &       7&0.243 & -4.67 &17. &9.23 &3.9 &7.0  \\ 
27905  &   0.627  &       3&0.150 & -5.06 &... &9.89 &2.3 &2.4  \\ 
28471  &   0.650  &       4&0.144 & -5.11 &... &9.91 &2.2 &2.3  \\ 
28701  &   0.650  &       4&0.145 & -5.10 &... &9.90 &2.2 &2.3  \\ 
28185  &   0.750  &      12&0.154 & -5.06 &... &9.88 &2.1 &2.1  \\ 
28821  &   0.683  &       5&0.158 & -5.02 &... &9.84 &2.4 &2.2  \\ 
29086  &   0.990  &      11&0.691 & ... &... &... &4.3 &7.7  \\ 
29220  &   1.106  &       5&0.787 & ... &... &... &1.6 &4.4  \\ 
29161  &   0.632  &       8&0.222 & -4.71 &15. &9.32 &3.6 &6.7  \\ 
30306  &   0.747  &       4&0.150 & -5.08 &... &9.91 &2.1 &2.1  \\ 
29813  &   0.621  &       4&0.153 & -5.03 &... &9.86 &2.4 &2.4  \\ 
30278  &   0.746  &      10&0.155 & -5.05 &... &9.88 &2.1 &2.1  \\ 
30669  &   0.795  &      12&0.159 & -5.04 &... &9.87 &2.1 &2.1  \\ 
33214  &   0.890  &       4&0.326 & -4.67 &27. &9.24 &2.8 &4.2  \\ 
hip22772  &   1.271  &       3&0.721 & ... &... &... &1.6 &2.1  \\ 
31392  &   0.792  &      10&0.270 & -4.69 &25. &9.27 &2.7 &4.9  \\ 
32564  &   0.741  &       5&0.150 & -5.08 &... &9.91 &2.1 &2.1  \\ 
32724  &   0.618  &      11&0.139 & -5.14 &... &9.96 &2.2 &2.5  \\ 
32842  &   0.492  &       7&0.179 & -4.83 &7. &9.57 &2.9 &6.3  \\ 
hip23708  &   1.386  &       1&1.737 & ... &... &... &4.7 &2.3  \\ 
35877  &   0.573  &       2&0.216 & -4.70 &10. &9.30 &3.5 &8.5  \\ 
33093  &   0.606  &      10&0.130 & -5.23 &... &10.07 &2.0 &2.6  \\ 
35267  &   0.513  &       5&0.129 & -5.24 &... &10.08 &2.1 &2.9  \\ 
33873  &   0.704  &       3&0.242 & -4.69 &20. &9.28 &2.5 &5.8  \\ 
33822  &   0.708  &       8&0.141 & -5.14 &... &9.95 &2.1 &2.1  \\ 
33693  &   0.472  &      15&0.137 & -5.14 &... &9.95 &2.1 &3.0  \\ 
hip24392  &   1.645  &       3&1.021 & ... &... &... &2.1 &2.5  \\ 
34195  &   0.567  &       8&0.128 & -5.26 &... &10.11 &2.0 &2.6  \\ 
35041  &   0.636  &       9&0.304 & -4.50 &8. &8.78 &5.0 &15.3  \\ 
36379  &   0.561  &       5&0.134 & -5.19 &... &10.02 &2.1 &2.7  \\ 
37351  &   0.594  &       5&0.191 & -4.81 &15. &9.52 &3.1 &5.0  \\ 
37761  &   0.764  &       4&0.125 & -5.27 &... &10.12 &2.0 &4.3  \\ 
37706  &   0.769  &       8&0.186 & -4.92 &36. &9.70 &2.3 &2.1  \\ 
37986  &   0.801  &       8&0.151 & -5.09 &... &9.91 &2.1 &2.1  \\ 
38459  &   0.861  &       1&0.471 & -4.46 &13. &8.66 &3.6 &9.3  \\ 
38683  &   0.571  &       3&0.180 & -4.85 &13. &9.60 &2.9 &3.8  \\ 
38277  &   0.637  &       7&0.139 & -5.15 &... &9.97 &2.1 &2.4  \\ 
38467  &   0.672  &       6&0.145 & -5.10 &... &9.90 &2.2 &2.2  \\ 
38554  &   0.704  &       8&0.175 & -4.93 &31. &9.73 &2.2 &2.1  \\ 
38677  &   0.581  &       6&0.137 & -5.16 &... &9.98 &2.1 &2.6  \\ 
hip27323  &   1.377  &       3&1.828 & ... &... &... &4.9 &2.2  \\ 
39503  &   0.732  &       5&0.160 & -5.02 &... &9.85 &2.1 &2.1  \\ 
hip27803  &   1.321  &       3&0.859 & ... &... &... &2.5 &2.1  \\ 
40105  &   0.891  &       8&0.137 & -5.17 &... &10.00 &2.0 &4.3  \\ 
39855  &   0.700  &       9&0.173 & -4.94 &31. &9.75 &2.7 &2.1  \\ 
hip28153  &   1.404  &       3&1.361 & ... &... &... &3.6 &2.3  \\ 
41158  &   0.519  &       2&0.142 & -5.09 &... &9.92 &2.3 &2.8  \\ 
41155  &   0.559  &       5&0.121 & -5.36 &... &10.22 &1.9 &2.6  \\ 
42044  &   0.543  &       5&0.117 & -5.43 &... &10.28 &1.8 &2.6  \\ 
42538  &   0.593  &       5&0.126 & -5.29 &... &10.14 &1.9 &2.6  \\ 
42719  &   0.657  &       5&0.128 & -5.27 &... &10.11 &1.9 &2.6  \\ 
43197  &   0.817  &       7&0.146 & -5.11 &... &9.91 &2.1 &2.1  \\ 
43470  &   0.539  &       2&0.134 & -5.18 &... &10.01 &2.1 &2.8  \\ 
43848  &   0.927  &       6&0.173 & ... &... &... &2.1 &2.1  \\ 
44310  &   0.837  &       2&0.156 & -5.07 &... &9.90 &2.1 &2.1  \\ 
44573  &   0.915  &       4&0.409 & ... &... &... &3.2 &5.4  \\ 
44569  &   0.619  &      10&0.157 & -5.01 &... &9.83 &2.5 &2.5  \\ 
45987  &   0.655  &      10&0.146 & -5.10 &... &9.92 &2.2 &2.3  \\ 
45738  &   0.460  &      11&0.163 & -4.93 &6. &9.72 &2.5 &3.6  \\ 
46435  &   0.702  &       2&0.224 & -4.74 &22. &9.39 &2.4 &4.9  \\ 
46894  &   0.778  &       5&0.136 & -5.17 &... &10.00 &2.0 &4.3  \\ 
47017  &   0.609  &       6&0.129 & -5.25 &... &10.09 &2.0 &2.5  \\ 
46872  &   0.536  &       2&0.205 & -4.73 &8. &9.36 &3.4 &8.8  \\ 
47855  &   0.465  &       7&0.189 & -4.79 &5. &9.48 &3.0 &8.2  \\ 
47186  &   0.714  &       8&0.150 & -5.08 &... &9.91 &2.1 &2.1  \\ 
hip31862  &   1.377  &       4&1.285 & ... &... &... &3.5 &2.2  \\ 
hip31878  &   1.257  &       9&2.339 & ... &... &... &1.6 &7.0  \\ 
48265  &   0.747  &      16&0.132 & -5.21 &... &10.04 &2.0 &4.3  \\ 
48056  &   0.576  &       2&0.160 & -4.97 &17. &9.78 &2.6 &2.6  \\ 
48286  &   0.563  &       7&0.153 & -5.01 &... &9.83 &2.5 &2.7  \\ 
49035  &   0.755  &       2&0.228 & -4.76 &27. &9.44 &2.5 &4.0  \\ 
hip32530  &   1.306  &       6&0.757 & ... &... &... &2.3 &2.1  \\ 
49866  &   0.663  &       2&0.157 & -5.02 &... &9.85 &2.4 &4.3  \\ 
51608  &   0.771  &       1&0.159 & -5.03 &... &9.86 &2.1 &2.1  \\ 
52217  &   0.823  &       8&0.229 & -4.81 &33. &9.52 &2.4 &2.1  \\ 
53466  &   0.551  &       1&0.133 & -5.20 &... &10.03 &2.1 &2.8  \\ 
53680  &   1.180  &       7&0.514 & ... &... &... &1.6 &2.1  \\ 
hip34785  &   1.330  &       4&0.811 & ... &... &... &2.3 &2.1  \\ 
56413  &   0.754  &       2&0.184 & -4.92 &35. &9.70 &2.3 &2.1  \\ 
56662  &   0.604  &      18&0.148 & -5.07 &... &9.90 &2.3 &2.5  \\ 
hip35173  &   1.010  &       5&0.320 & ... &... &... &1.6 &2.1  \\ 
57553  &   0.506  &       1&0.172 & -4.88 &8. &9.64 &2.8 &2.8  \\ 
hip35937  &   1.274  &       2&0.815 & ... &... &... &1.6 &2.1  \\ 
hip35943  &   1.365  &       3&1.741 & ... &... &... &4.7 &2.2  \\ 
58556  &   0.568  &      14&0.204 & -4.74 &11. &9.39 &3.3 &7.2  \\ 
59711  &   0.637  &       6&0.156 & -5.02 &... &9.84 &2.4 &2.4  \\ 
61475  &   0.853  &       7&0.340 & -4.61 &23. &9.10 &2.9 &5.4  \\ 
61214  &   1.049  &       8&0.815 & ... &... &... &1.6 &6.7  \\ 
61051  &   0.766  &       3&0.158 & -5.04 &... &9.86 &2.1 &2.1  \\ 
60779  &   0.564  &      10&0.149 & -5.05 &... &9.87 &2.4 &2.7  \\ 
hip37727  &   0.700  &      15&0.296 & -4.56 &14. &8.95 &4.8 &9.7  \\ 
62549  &   0.612  &      10&0.150 & -5.05 &... &9.88 &2.3 &2.5  \\ 
64114  &   0.721  &       5&0.218 & -4.77 &25. &9.45 &2.4 &4.2  \\ 
hip38594  &   1.385  &       7&1.087 & ... &... &... &2.9 &2.3  \\ 
hip38910  &   1.136  &       8&0.621 & ... &... &... &1.6 &2.1  \\ 
66039  &   0.582  &       2&0.199 & -4.77 &13. &9.45 &3.2 &6.1  \\ 
67200  &   0.595  &       1&0.131 & -5.22 &... &10.06 &2.0 &2.6  \\ 
68475  &   0.887  &       6&0.305 & -4.71 &29. &9.31 &2.7 &3.8  \\ 
68607  &   0.862  &       1&0.277 & -4.73 &30. &9.37 &2.6 &3.6  \\ 
69611  &   0.584  &       8&0.147 & -5.06 &... &9.89 &2.3 &2.6  \\ 
70081  &   0.660  &       2&0.140 & -5.14 &... &9.96 &2.1 &2.3  \\ 
72579  &   0.790  &       8&0.157 & -5.05 &... &9.88 &2.1 &2.1  \\ 
73744  &   0.607  &       5&0.157 & -5.00 &21. &9.82 &2.5 &2.5  \\ 
73267  &   0.806  &       3&0.154 & -5.07 &... &9.90 &2.1 &2.1  \\ 
73322  &   0.910  &      22&0.458 & ... &... &... &3.4 &6.6  \\ 
hip42507  &   1.361  &       2&1.654 & ... &... &... &4.5 &2.2  \\ 
hip42601  &   1.360  &       6&0.994 & ... &... &... &2.7 &2.2  \\ 
hip42881  &   1.432  &       5&1.153 & ... &... &... &3.1 &2.3  \\ 
75351  &   0.625  &       3&0.142 & -5.13 &... &9.93 &2.2 &2.6  \\ 
75288  &   0.673  &       2&0.149 & -5.08 &... &9.90 &2.3 &2.2  \\ 
77417  &   0.770  &      16&0.177 & -4.95 &38. &9.75 &2.2 &2.1  \\ 
77825  &   0.960  &      21&0.755 & ... &... &... &4.7 &10.8  \\ 
hip44722  &   1.418  &       7&1.937 & ... &... &... &5.2 &2.3  \\ 
78747  &   0.575  &      22&0.146 & -5.07 &... &9.90 &2.3 &2.6  \\ 
78558  &   0.617  &      17&0.151 & -5.05 &... &9.88 &2.4 &2.5  \\ 
78612  &   0.610  &      22&0.142 & -5.11 &... &9.92 &2.2 &2.5  \\ 
hip44899  &   1.316  &      10&1.007 & ... &... &... &2.9 &2.1  \\ 
79985  &   0.659  &       3&0.126 & -5.28 &... &10.13 &1.9 &2.6  \\ 
81238  &   0.509  &       1&0.157 & -4.97 &10. &9.78 &2.5 &2.9  \\ 
81110  &   0.724  &      15&0.193 & -4.87 &29. &9.62 &2.3 &2.1  \\ 
81659  &   0.670  &      18&0.251 & -4.64 &15. &9.17 &4.1 &7.7  \\ 
hip46549  &   1.305  &       8&1.021 & ... &... &... &3.0 &2.1  \\ 
82342  &   0.985  &       9&0.272 & ... &... &... &2.4 &2.1  \\ 
82516  &   0.927  &       9&0.228 & ... &... &... &2.3 &2.1  \\ 
85249  &   0.492  &       1&0.132 & -5.19 &... &10.03 &2.1 &3.0  \\ 
85390  &   0.855  &      10&0.210 & -4.89 &40. &9.66 &2.3 &2.1  \\ 
hip48336  &   1.446  &       7&0.764 & ... &... &... &2.0 &2.4  \\ 
85380  &   0.577  &      10&0.141 & -5.12 &... &9.92 &2.2 &2.6  \\ 
hip48502  &   1.375  &       3&0.835 & ... &... &... &2.3 &2.2  \\ 
86006  &   0.708  &       2&0.140 & -5.15 &... &9.97 &2.0 &2.6  \\ 
86249  &   0.943  &       4&0.341 & ... &... &... &2.8 &2.1  \\ 
86226  &   0.647  &      15&0.168 & -4.95 &25. &9.75 &2.6 &2.3  \\ 
86397  &   0.725  &       7&0.175 & -4.94 &33. &9.74 &2.2 &2.1  \\ 
86652  &   0.630  &       2&0.153 & -5.04 &... &9.86 &2.4 &2.4  \\ 
87966  &   0.415  &       1&0.168 & -4.90 &3. &9.68 &2.3 &3.0  \\ 
hip49577  &   1.331  &       4&0.936 & ... &... &... &2.7 &2.1  \\ 
87931  &   0.760  &       3&0.161 & -5.02 &... &9.85 &2.1 &2.1  \\ 
87998  &   0.622  &      21&0.147 & -5.08 &... &9.91 &2.3 &2.4  \\ 
89591  &   0.642  &       5&0.165 & -4.96 &25. &9.77 &2.6 &2.6  \\ 
89418  &   0.569  &      20&0.156 & -4.99 &16. &9.81 &2.5 &2.7  \\ 
89839  &   0.523  &       3&0.130 & -5.22 &... &10.06 &2.1 &2.9  \\ 
89942  &   0.700  &       1&0.144 & -5.11 &... &9.92 &2.2 &2.1  \\ 
89988  &   0.681  &       2&0.141 & -5.14 &... &9.95 &2.1 &2.6  \\ 
90028  &   0.656  &      12&0.154 & -5.04 &... &9.86 &2.4 &2.3  \\ 
90926  &   0.748  &       2&0.158 & -5.03 &... &9.86 &2.1 &2.1  \\ 
90520  &   0.643  &       1&0.141 & -5.14 &... &9.95 &2.2 &2.6  \\ 
90884  &   1.045  &      14&0.830 & ... &... &... &1.6 &7.0  \\ 
91682  &   0.696  &       1&0.145 & -5.11 &... &9.91 &2.2 &2.3  \\ 
91320  &   0.588  &       1&0.137 & -5.16 &... &9.98 &2.1 &2.6  \\ 
91901  &   0.920  &      11&0.420 & ... &... &... &3.2 &5.5  \\ 
94009  &   0.770  &       1&0.134 & -5.19 &... &10.02 &2.0 &4.3  \\ 
hip52296  &   1.464  &       6&1.167 & ... &... &... &3.2 &2.4  \\ 
hip52341  &   1.460  &       2&1.062 & ... &... &... &2.9 &2.4  \\ 
92719  &   0.622  &      13&0.179 & -4.88 &19. &9.64 &2.8 &2.4  \\ 
93083  &   0.945  &      10&0.185 & ... &... &... &2.1 &2.1  \\ 
93489  &   0.620  &      12&0.155 & -5.02 &... &9.85 &2.4 &2.4  \\ 
94151  &   0.718  &      14&0.172 & -4.95 &33. &9.76 &2.2 &2.1  \\ 
94270  &   0.579  &      11&0.149 & -5.05 &... &9.87 &2.4 &2.6  \\ 
94482  &   0.562  &      15&0.138 & -5.14 &... &9.96 &2.2 &2.7  \\ 
94690  &   0.710  &       2&0.169 & -4.96 &33. &9.77 &2.2 &2.1  \\ 
94838  &   0.635  &      10&0.214 & -4.74 &16. &9.38 &3.4 &5.9  \\ 
95090  &   0.639  &       1&0.141 & -5.13 &... &9.94 &2.2 &2.4  \\ 
95136  &   0.660  &       6&0.151 & -5.06 &... &9.89 &2.3 &2.3  \\ 
95338  &   0.878  &       8&0.195 & -4.94 &44. &9.75 &2.2 &2.1  \\ 
95521  &   0.637  &      16&0.163 & -4.97 &24. &9.79 &2.5 &2.4  \\ 
96417  &   0.563  &       1&0.135 & -5.17 &... &10.00 &2.1 &2.7  \\ 
hip54373  &   1.430  &       8&1.762 & ... &... &... &4.7 &2.3  \\ 
hip54569  &   1.400  &       5&1.260 & ... &... &... &3.4 &2.3  \\ 
97782  &   1.108  &      12&0.452 & ... &... &... &1.6 &2.1  \\ 
hip54966  &   1.340  &       7&1.003 & ... &... &... &2.8 &2.2  \\ 
98179  &   0.500  &       1&0.140 & -5.11 &... &9.91 &2.2 &3.0  \\ 
hip55119  &   1.383  &       7&1.540 & ... &... &... &4.1 &2.3  \\ 
98459  &   0.623  &       1&0.130 & -5.24 &... &10.09 &2.0 &2.4  \\ 
98640  &   0.765  &       1&0.338 & -4.53 &15. &8.88 &3.0 &2.6  \\ 
98553  &   0.594  &      17&0.154 & -5.02 &... &9.84 &2.4 &2.6  \\ 
98649  &   0.658  &       5&0.154 & -5.04 &... &9.86 &2.4 &2.3  \\ 
98727  &   0.521  &       1&0.166 & -4.91 &10. &9.70 &2.7 &4.0  \\ 
100508  &   0.828  &       9&0.158 & -5.06 &... &9.88 &2.1 &2.1  \\ 
100850  &   0.661  &       3&0.135 & -5.20 &... &10.03 &2.0 &2.3  \\ 
101117  &   0.638  &       1&0.251 & -4.63 &13. &9.13 &4.1 &9.1  \\ 
101093  &   0.561  &       7&0.161 & -4.96 &15. &9.77 &2.6 &2.7  \\ 
101171  &   0.771  &       4&0.193 & -4.89 &35. &9.66 &2.3 &2.1  \\ 
101197  &   0.659  &       1&0.152 & -5.05 &... &9.88 &2.3 &2.3  \\ 
101930  &   0.908  &      13&0.188 & ... &... &... &2.2 &2.1  \\ 
102574  &   0.585  &      13&0.130 & -5.23 &... &10.07 &2.0 &2.6  \\ 
hip57688  &   1.078  &       8&0.700 & ... &... &... &1.6 &4.5  \\ 
102804  &   0.413  &       1&0.202 & -4.73 &2. &9.37 &2.9 &10.2  \\ 
102843  &   0.787  &       7&0.186 & -4.92 &38. &9.71 &2.3 &2.1  \\ 
103220  &   0.740  &       2&0.151 & -5.07 &... &9.90 &2.1 &2.6  \\ 
103743  &   0.640  &       1&0.351 & -4.42 &6. &8.53 &5.8 &21.3  \\ 
103760  &   0.650  &      19&0.148 & -5.08 &... &9.91 &2.3 &2.3  \\ 
103836  &   1.024  &      11&0.474 & ... &... &... &1.6 &3.6  \\ 
103949  &   0.985  &       8&0.350 & ... &... &... &2.7 &2.1  \\ 
hip58688  &   1.419  &       6&1.182 & ... &... &... &3.1 &2.3  \\ 
104760  &   0.645  &       4&0.157 & -5.01 &... &9.83 &2.4 &2.3  \\ 
104982  &   0.651  &      14&0.166 & -4.96 &26. &9.77 &2.6 &2.3  \\ 
105671  &   1.137  &      10&0.869 & ... &... &... &1.6 &4.2  \\ 
105837  &   0.570  &      13&0.159 & -4.97 &16. &9.79 &2.5 &2.7  \\ 
105904  &   0.742  &       6&0.143 & -5.13 &... &9.94 &2.1 &2.1  \\ 
106275  &   0.892  &       7&0.202 & -4.94 &44. &9.73 &2.3 &2.1  \\ 
106290  &   0.623  &       1&0.170 & -4.93 &21. &9.72 &2.7 &2.4  \\ 
106515A  &   0.815  &       7&0.156 & -5.06 &... &9.89 &2.1 &2.6  \\ 
106515B  &   0.815  &       8&0.161 & -5.04 &... &9.87 &2.1 &2.1  \\ 
106869  &   0.574  &       8&0.140 & -5.12 &... &9.93 &2.2 &2.7  \\ 
106937  &   0.762  &       8&0.141 & -5.14 &... &9.95 &2.1 &2.6  \\ 
107008  &   0.841  &       7&0.257 & -4.76 &31. &9.42 &2.6 &2.1  \\ 
107077  &   0.435  &       7&0.165 & -4.92 &4. &9.70 &2.4 &3.0  \\ 
107352  &   0.571  &       1&0.212 & -4.71 &10. &9.33 &3.5 &8.1  \\ 
107388  &   1.009  &      14&0.549 & ... &... &... &1.6 &4.8  \\ 
107576  &   1.064  &      12&0.440 & ... &... &... &1.6 &2.1  \\ 
107820  &   0.863  &       1&0.140 & -5.16 &... &9.97 &2.0 &4.3  \\ 
107956  &   0.475  &       1&0.202 & -4.73 &5. &9.36 &3.3 &10.9  \\ 
108446  &   0.872  &      11&0.302 & -4.70 &28. &9.29 &2.7 &4.0  \\ 
108953  &   0.800  &      11&0.179 & -4.96 &41. &9.76 &2.2 &2.1  \\ 
109067  &   0.405  &       1&0.164 & -4.93 &3. &9.72 &2.2 &3.0  \\ 
109591  &   0.666  &      12&0.148 & -5.09 &... &9.91 &2.3 &2.2  \\ 
hip61629  &   1.470  &      11&1.705 & ... &... &... &4.6 &2.4  \\ 
bd-043319b  &   0.980  &       2&0.308 & ... &... &... &2.5 &2.1  \\ 
109930  &   0.894  &       7&0.202 & -4.94 &45. &9.74 &2.3 &2.1  \\ 
109908  &   0.583  &      11&0.268 & -4.55 &7. &8.93 &4.4 &15.3  \\ 
109952  &   1.154  &       5&0.623 & ... &... &... &1.6 &2.1  \\ 
110273  &   0.538  &       2&0.142 & -5.10 &... &9.90 &2.3 &2.8  \\ 
110605  &   0.730  &       8&0.171 & -4.96 &35. &9.77 &2.2 &2.1  \\ 
110619  &   0.664  &      14&0.169 & -4.95 &27. &9.75 &2.6 &2.2  \\ 
111261B  &   1.380  &       2&0.881 & ... &... &... &2.4 &2.2  \\ 
111261A  &   1.380  &       6&0.779 & ... &... &... &2.1 &2.2  \\ 
111232  &   0.701  &      18&0.160 & -5.02 &... &9.84 &2.1 &2.1  \\ 
111431  &   0.627  &      11&0.131 & -5.23 &... &10.07 &2.0 &2.6  \\ 
112121  &   0.728  &       4&0.149 & -5.09 &... &9.91 &2.1 &2.1  \\ 
112456  &   0.569  &       4&0.160 & -4.97 &16. &9.78 &2.6 &2.7  \\ 
113027  &   0.569  &      13&0.187 & -4.82 &12. &9.53 &3.0 &5.2  \\ 
113449  &   0.847  &       4&0.581 & -4.34 &6. &8.27 &4.1 &14.9  \\ 
113538  &   1.362  &      11&1.160 & ... &... &... &3.2 &2.2  \\ 
113693  &   1.035  &       5&0.273 & ... &... &... &1.6 &2.1  \\ 
114386  &   0.982  &       7&0.244 & ... &... &... &2.2 &2.1  \\ 
114432  &   0.756  &       9&0.357 & -4.50 &13. &8.77 &3.1 &10.5  \\ 
114747  &   0.923  &       5&0.283 & ... &... &... &2.6 &2.1  \\ 
115080  &   0.683  &       5&0.156 & -5.03 &... &9.86 &2.4 &2.2  \\ 
115106  &   0.660  &       1&0.134 & -5.20 &... &10.04 &2.0 &2.6  \\ 
116220  &   0.667  &       1&0.138 & -5.16 &... &9.99 &2.1 &2.2  \\ 
116259  &   0.721  &       1&0.141 & -5.14 &... &9.95 &2.1 &2.1  \\ 
115674  &   0.667  &      13&0.186 & -4.86 &24. &9.62 &2.9 &2.2  \\ 
116920  &   0.911  &       7&0.251 & ... &... &... &2.5 &2.1  \\ 
117860  &   0.624  &      12&0.281 & -4.54 &9. &8.90 &4.6 &13.4  \\ 
118926  &   1.376  &       3&0.928 & ... &... &... &2.5 &2.2  \\ 
hip66678  &   1.337  &      16&0.803 & ... &... &... &2.3 &2.2  \\ 
119217  &   1.288  &       3&0.910 & ... &... &... &1.6 &2.1  \\ 
119119  &   0.589  &       2&0.126 & -5.29 &... &10.14 &1.9 &2.6  \\ 
119329  &   0.735  &       1&0.147 & -5.10 &... &9.90 &2.1 &2.5  \\ 
119782  &   0.856  &       9&0.305 & -4.67 &27. &9.24 &2.8 &4.4  \\ 
120036B  &   1.322  &       4&1.367 & ... &... &... &3.9 &2.1  \\ 
120036A  &   1.322  &       8&1.525 & ... &... &... &4.3 &2.1  \\ 
120056  &   0.724  &       3&0.135 & -5.19 &... &10.02 &2.0 &4.3  \\ 
120250  &   0.461  &       5&0.126 & -5.27 &... &10.12 &1.9 &3.0  \\ 
120329  &   0.739  &       5&0.143 & -5.13 &... &9.94 &2.1 &2.6  \\ 
120744  &   0.913  &       7&0.205 & ... &... &... &2.2 &2.1  \\ 
121504  &   0.593  &      20&0.175 & -4.89 &17. &9.66 &2.8 &3.6  \\ 
122245  &   0.485  &       1&0.164 & -4.92 &7. &9.71 &2.6 &4.1  \\ 
124227  &   0.529  &       1&0.172 & -4.88 &10. &9.64 &2.8 &4.6  \\ 
hip69454  &   1.500  &       5&0.992 & ... &... &... &2.9 &2.5  \\ 
124553  &   0.593  &       8&0.159 & -4.98 &19. &9.80 &2.5 &2.6  \\ 
124595  &   0.610  &       5&0.149 & -5.06 &... &9.89 &2.3 &2.5  \\ 
124925  &   0.920  &       2&0.155 & ... &... &... &2.0 &4.3  \\ 
125595  &   1.107  &       5&0.537 & ... &... &... &1.6 &2.1  \\ 
125906B  &   0.584  &       8&0.155 & -5.01 &... &9.83 &2.5 &2.6  \\ 
125906A  &   0.584  &      14&0.156 & -5.00 &18. &9.82 &2.5 &2.6  \\ 
126525  &   0.682  &       9&0.157 & -5.03 &... &9.85 &2.4 &2.2  \\ 
126653  &   0.574  &       2&0.164 & -4.94 &16. &9.74 &2.6 &2.6  \\ 
127339  &   1.403  &       5&1.144 & ... &... &... &3.1 &2.3  \\ 
127321  &   0.615  &       1&0.130 & -5.24 &... &10.09 &2.0 &2.5  \\ 
127423  &   0.569  &      13&0.225 & -4.67 &9. &9.23 &3.7 &9.8  \\ 
hip71135  &   1.270  &       8&0.783 & ... &... &... &1.6 &2.1  \\ 
128214  &   0.720  &       2&0.173 & -4.95 &33. &9.76 &2.2 &2.6  \\ 
128760  &   0.568  &       4&0.158 & -4.98 &16. &9.79 &2.5 &2.7  \\ 
128987  &   0.710  &      13&0.360 & -4.45 &9. &8.64 &3.1 &14.2  \\ 
129679  &   0.600  &       1&0.315 & -4.46 &6. &8.66 &5.2 &21.1  \\ 
129642  &   0.936  &       9&0.213 & ... &... &... &2.2 &2.1  \\ 
129903  &   0.576  &       1&0.386 & -4.32 &2. &8.22 &6.5 &43.3  \\ 
129445  &   0.756  &      12&0.161 & -5.02 &... &9.84 &2.1 &2.1  \\ 
129946  &   0.710  &       1&0.280 & -4.60 &16. &9.06 &2.7 &8.0  \\ 
130102  &   0.544  &       1&0.178 & -4.86 &11. &9.61 &2.9 &4.8  \\ 
130930  &   0.913  &       6&0.190 & ... &... &... &2.2 &2.1  \\ 
131719  &   1.005  &      19&0.607 & ... &... &... &1.6 &5.7  \\ 
131900  &   0.747  &       1&0.170 & -4.97 &37. &9.79 &2.2 &2.1  \\ 
hip73362  &   1.318  &       1&1.242 & ... &... &... &3.5 &2.1  \\ 
131664  &   0.667  &       8&0.179 & -4.90 &25. &9.67 &2.8 &2.2  \\ 
hip73427  &   1.253  &       4&1.202 & ... &... &... &1.6 &2.1  \\ 
132648  &   0.721  &       9&0.191 & -4.87 &29. &9.63 &2.3 &2.1  \\ 
132899  &   1.145  &       6&1.032 & ... &... &... &1.6 &5.0  \\ 
132996  &   0.613  &      15&0.154 & -5.02 &... &9.84 &2.4 &2.5  \\ 
133131A  &   0.622  &      13&0.160 & -4.99 &23. &9.80 &2.5 &2.4  \\ 
133131B  &   0.622  &       2&0.164 & -4.96 &22. &9.77 &2.6 &2.4  \\ 
134048  &   0.727  &       3&0.139 & -5.16 &... &9.98 &2.0 &2.6  \\ 
133639  &   0.678  &       5&0.163 & -4.99 &30. &9.81 &2.5 &2.2  \\ 
134664  &   0.662  &       9&0.172 & -4.93 &26. &9.73 &2.7 &2.3  \\ 
134928  &   0.759  &       2&0.212 & -4.82 &30. &9.54 &2.4 &2.1  \\ 
134929  &   0.810  &       1&0.193 & -4.91 &39. &9.69 &2.3 &2.1  \\ 
135446  &   0.649  &       7&0.124 & -5.31 &... &10.16 &1.9 &2.6  \\ 
135309  &   0.640  &       3&0.170 & -4.94 &23. &9.73 &2.7 &2.4  \\ 
135725  &   0.748  &       6&0.166 & -4.99 &38. &9.81 &2.2 &2.1  \\ 
135562  &   0.652  &       1&0.129 & -5.25 &... &10.09 &1.9 &2.6  \\ 
135625  &   0.617  &       3&0.146 & -5.09 &... &9.91 &2.3 &2.5  \\ 
136762  &   0.624  &       3&0.236 & -4.66 &13. &9.21 &3.8 &2.5  \\ 
136894  &   0.721  &       8&0.173 & -4.95 &33. &9.75 &2.2 &2.1  \\ 
137214  &   0.593  &       2&0.149 & -5.06 &... &9.88 &2.3 &2.6  \\ 
137628  &   1.010  &       5&0.348 & ... &... &... &1.6 &2.1  \\ 
137388  &   0.891  &       6&0.265 & -4.78 &35. &9.47 &2.6 &2.1  \\ 
139061  &   0.412  &       1&0.148 & -5.04 &... &9.87 &2.0 &3.0  \\ 
139763  &   1.296  &       6&1.405 & ... &... &... &1.6 &2.1  \\ 
140643  &   0.898  &       2&0.293 & -4.74 &32. &9.38 &2.7 &2.1  \\ 
141514  &   0.588  &      10&0.152 & -5.03 &... &9.86 &2.4 &2.6  \\ 
141366  &   0.683  &       2&0.142 & -5.13 &... &9.94 &2.2 &4.3  \\ 
141599  &   0.757  &       5&0.209 & -4.82 &30. &9.55 &2.4 &2.1  \\ 
141815  &   0.542  &       2&0.129 & -5.24 &... &10.08 &2.1 &2.8  \\ 
142072  &   0.670  &      12&0.310 & -4.51 &10. &8.81 &5.1 &13.1  \\ 
141885  &   0.652  &       2&0.122 & -5.34 &... &10.19 &1.8 &4.3  \\ 
142137  &   0.629  &       3&0.142 & -5.12 &... &9.92 &2.2 &4.2  \\ 
142709  &   1.118  &       8&0.419 & ... &... &... &1.6 &2.1  \\ 
hip78353  &   1.497  &       1&1.353 & ... &... &... &3.8 &2.5  \\ 
143120  &   0.780  &       3&0.131 & -5.21 &... &10.04 &2.0 &4.3  \\ 
143137  &   0.518  &       2&0.129 & -5.23 &... &10.07 &2.1 &2.9  \\ 
143361  &   0.773  &      13&0.167 & -5.00 &41. &9.82 &2.2 &2.1  \\ 
143673  &   0.802  &       4&0.163 & -5.02 &... &9.85 &2.1 &2.2  \\ 
143846  &   0.600  &       3&0.165 & -4.94 &19. &9.75 &2.6 &2.5  \\ 
144087  &   0.750  &       6&0.313 & -4.57 &16. &8.97 &2.9 &8.3  \\ 
144088  &   0.850  &       4&0.354 & -4.59 &21. &9.03 &3.0 &6.0  \\ 
144550  &   0.682  &       4&0.159 & -5.01 &... &9.84 &2.4 &2.2  \\ 
144766  &   0.562  &      12&0.157 & -4.98 &15. &9.80 &2.5 &2.7  \\ 
144848  &   0.648  &       2&0.152 & -5.05 &... &9.87 &2.4 &2.3  \\ 
142022  &   0.790  &       6&0.192 & -4.90 &37. &9.68 &2.3 &2.1  \\ 
144899  &   0.660  &       4&0.150 & -5.07 &... &9.89 &2.3 &2.3  \\ 
145518  &   0.616  &       6&0.209 & -4.75 &15. &9.40 &3.4 &6.0  \\ 
145666  &   0.603  &      11&0.193 & -4.80 &15. &9.51 &3.1 &4.9  \\ 
146070  &   0.614  &       2&0.212 & -4.73 &14. &9.37 &3.4 &6.3  \\ 
146124  &   0.760  &       6&0.293 & -4.61 &19. &9.09 &2.8 &6.8  \\ 
146835  &   0.585  &       3&0.199 & -4.77 &13. &9.45 &3.2 &6.0  \\ 
146817  &   0.666  &      10&0.149 & -5.07 &... &9.90 &2.3 &2.2  \\ 
hip80018  &   1.550  &       5&0.680 & ... &... &... &2.4 &2.5  \\ 
147018  &   0.763  &       9&0.219 & -4.80 &29. &9.50 &2.4 &3.6  \\ 
144167  &   0.651  &       4&0.191 & -4.84 &21. &9.57 &3.0 &2.3  \\ 
147873  &   0.575  &       4&0.128 & -5.26 &... &10.10 &2.0 &2.6  \\ 
147619  &   0.604  &       1&0.125 & -5.31 &... &10.16 &1.9 &2.5  \\ 
148156  &   0.560  &       2&0.151 & -5.03 &... &9.85 &2.4 &2.7  \\ 
148303  &   0.980  &       1&0.465 & ... &... &... &3.3 &4.5  \\ 
148290  &   0.502  &       1&0.148 & -5.03 &... &9.86 &2.4 &3.0  \\ 
148628  &   0.538  &       3&0.139 & -5.13 &... &9.94 &2.2 &2.8  \\ 
148587  &   0.573  &      10&0.139 & -5.13 &... &9.95 &2.2 &2.7  \\ 
149189  &   0.653  &       1&0.136 & -5.18 &... &10.01 &2.1 &2.3  \\ 
149192  &   1.096  &       3&0.799 & ... &... &... &1.6 &4.8  \\ 
149396  &   0.704  &       6&0.257 & -4.65 &18. &9.19 &2.6 &6.8  \\ 
149079  &   0.576  &      10&0.142 & -5.11 &... &9.90 &2.3 &2.6  \\ 
149194  &   0.559  &       6&0.151 & -5.03 &... &9.86 &2.4 &2.7  \\ 
149606  &   0.965  &       3&0.227 & ... &... &... &2.2 &2.1  \\ 
151450  &   0.566  &       3&0.154 & -5.01 &... &9.83 &2.5 &2.7  \\ 
hip82357  &   1.330  &       4&1.509 & ... &... &... &4.2 &2.1  \\ 
150761  &   0.564  &       1&0.145 & -5.08 &... &9.90 &2.3 &2.7  \\ 
152079  &   0.711  &      11&0.164 & -4.99 &34. &9.81 &2.2 &2.1  \\ 
152138  &   0.490  &      10&0.202 & -4.73 &5. &9.36 &3.3 &10.5  \\ 
152388  &   0.904  &       4&0.364 & ... &... &... &3.0 &4.7  \\ 
153026  &   1.181  &       6&0.308 & ... &... &... &1.6 &2.1  \\ 
154518  &   1.015  &       5&0.442 & ... &... &... &1.6 &2.1  \\ 
154697  &   0.735  &       2&0.222 & -4.77 &26. &9.45 &2.4 &4.1  \\ 
154682  &   0.617  &       7&0.143 & -5.11 &... &9.91 &2.2 &2.5  \\ 
154672  &   0.713  &      14&0.156 & -5.04 &... &9.87 &2.1 &2.1  \\ 
hip84452  &   1.273  &       2&1.498 & ... &... &... &1.6 &3.6  \\ 
156152  &   0.644  &       6&0.143 & -5.12 &... &9.92 &2.2 &2.3  \\ 
156239  &   0.942  &       2&0.118 & ... &... &... &1.8 &4.3  \\ 
156643  &   0.627  &       5&0.161 & -4.98 &23. &9.79 &2.5 &2.4  \\ 
157830  &   0.685  &       7&0.195 & -4.84 &24. &9.57 &3.1 &3.6  \\ 
hip85523  &   1.553  &       4&1.530 & ... &... &... &4.8 &2.5  \\ 
158469  &   0.556  &       6&0.133 & -5.19 &... &10.02 &2.1 &2.7  \\ 
158630  &   0.600  &       9&0.157 & -5.00 &... &9.82 &2.5 &2.5  \\ 
159902  &   0.682  &       1&0.149 & -5.08 &... &9.90 &2.3 &2.2  \\ 
160411  &   0.655  &       5&0.122 & -5.34 &... &10.19 &1.8 &3.9  \\ 
161098  &   0.676  &       7&0.163 & -4.99 &30. &9.81 &2.5 &2.2  \\ 
160859  &   0.617  &      11&0.170 & -4.92 &20. &9.72 &2.7 &2.5  \\ 
162907  &   0.758  &      12&0.159 & -5.03 &... &9.86 &2.1 &2.1  \\ 
163568  &   0.556  &       4&0.152 & -5.02 &... &9.85 &2.4 &2.7  \\ 
164604  &   1.396  &      15&0.547 & ... &... &... &1.4 &2.3  \\ 
165011  &   0.576  &       3&0.141 & -5.12 &... &9.93 &2.2 &2.6  \\ 
165204  &   0.770  &       5&0.160 & -5.03 &... &9.86 &2.1 &2.1  \\ 
165385  &   0.581  &       5&0.130 & -5.24 &... &10.08 &2.0 &2.6  \\ 
165271  &   0.655  &       5&0.137 & -5.17 &... &10.00 &2.1 &2.6  \\ 
165920  &   0.900  &       3&0.152 & -5.11 &... &9.91 &2.0 &2.1  \\ 
166184  &   1.020  &       2&0.578 & ... &... &... &1.6 &4.9  \\ 
165499  &   0.592  &       4&0.156 & -5.00 &19. &9.82 &2.5 &2.6  \\ 
166348  &   1.297  &       4&1.180 & ... &... &... &1.6 &2.1  \\ 
166745  &   0.763  &       4&0.155 & -5.06 &... &9.88 &2.1 &2.1  \\ 
168788  &   0.845  &       4&0.263 & -4.75 &31. &9.40 &2.6 &3.6  \\ 
169303  &   0.564  &       4&0.130 & -5.23 &... &10.07 &2.0 &2.6  \\ 
170641  &   0.438  &      13&0.148 & -5.03 &... &9.86 &2.2 &3.0  \\ 
169506  &   0.608  &       3&0.130 & -5.24 &... &10.09 &2.0 &3.0  \\ 
172063  &   0.756  &       6&0.179 & -4.93 &36. &9.73 &2.2 &2.1  \\ 
172582  &   0.685  &       9&0.160 & -5.01 &... &9.83 &2.5 &2.2  \\ 
173206  &   0.619  &       4&0.155 & -5.02 &... &9.84 &2.4 &2.5  \\ 
171990  &   0.593  &      13&0.139 & -5.14 &... &9.95 &2.2 &2.6  \\ 
173872  &   0.889  &       3&0.198 & -4.94 &45. &9.74 &2.2 &2.1  \\ 
174494  &   0.694  &       4&0.144 & -5.12 &... &9.92 &2.2 &2.6  \\ 
174541  &   0.631  &       4&0.143 & -5.12 &... &9.92 &2.2 &2.4  \\ 
174545  &   0.884  &       5&0.194 & -4.95 &45. &9.76 &2.2 &2.1  \\ 
175073  &   0.857  &       2&0.323 & -4.64 &25. &9.17 &2.9 &4.9  \\ 
175224  &   1.433  &      11&1.858 & ... &... &... &5.0 &2.4  \\ 
175167  &   0.751  &       9&0.152 & -5.07 &... &9.90 &2.1 &2.6  \\ 
176151  &   0.698  &       4&0.140 & -5.15 &... &9.96 &2.1 &2.6  \\ 
176110  &   0.698  &      16&0.144 & -5.12 &... &9.92 &2.2 &3.2  \\ 
176986  &   0.939  &       5&0.275 & ... &... &... &2.5 &2.1  \\ 
176354  &   0.888  &       7&0.178 & -5.01 &... &9.83 &2.1 &4.3  \\ 
177122  &   0.591  &       3&0.151 & -5.04 &... &9.86 &2.4 &2.6  \\ 
175169  &   0.721  &       4&0.159 & -5.02 &... &9.85 &2.1 &2.1  \\ 
178076  &   0.787  &       9&0.398 & -4.47 &12. &8.68 &3.3 &10.8  \\ 
178340  &   0.757  &       4&0.163 & -5.01 &... &9.83 &2.2 &2.1  \\ 
177409  &   0.600  &       7&0.166 & -4.94 &19. &9.74 &2.6 &2.5  \\ 
179205  &   0.579  &       4&0.266 & -4.55 &7. &8.94 &4.4 &15.4  \\ 
179640  &   0.738  &       7&0.300 & -4.58 &17. &9.01 &2.8 &8.1  \\ 
179699  &   0.596  &       4&0.131 & -5.22 &... &10.06 &2.0 &2.6  \\ 
hip94674  &   1.420  &       4&0.733 & ... &... &... &1.9 &2.3  \\ 
180409  &   0.570  &      11&0.150 & -5.04 &... &9.86 &2.4 &2.7  \\ 
180204  &   0.607  &       3&0.145 & -5.09 &... &9.92 &2.3 &2.5  \\ 
180257  &   0.644  &       1&0.130 & -5.24 &... &10.08 &2.0 &2.5  \\ 
181010  &   0.842  &       3&0.207 & -4.88 &39. &9.65 &2.3 &2.1  \\ 
182228  &   0.630  &       3&0.148 & -5.08 &... &9.90 &2.3 &2.4  \\ 
181433  &   1.006  &       6&0.180 & ... &... &... &1.6 &2.1  \\ 
182498  &   0.601  &       4&0.165 & -4.95 &19. &9.75 &2.6 &2.5  \\ 
183783  &   1.088  &       4&0.398 & ... &... &... &1.6 &2.1  \\ 
183804  &   0.564  &       4&0.273 & -4.53 &6. &8.87 &4.5 &18.2  \\ 
184081  &   0.757  &       2&0.165 & -5.00 &... &9.82 &2.2 &3.4  \\ 
184317  &   0.614  &      10&0.153 & -5.03 &... &9.86 &2.4 &2.5  \\ 
185615  &   0.734  &       3&0.156 & -5.04 &... &9.87 &2.1 &2.1  \\ 
185283  &   1.040  &       6&0.230 & ... &... &... &1.6 &2.1  \\ 
hip96998  &   1.347  &      12&0.903 & ... &... &... &2.5 &2.2  \\ 
185928  &   0.947  &       4&0.128 & ... &... &... &1.8 &4.3  \\ 
186265  &   0.790  &       5&0.146 & -5.11 &... &9.91 &2.1 &2.1  \\ 
186061  &   0.990  &       7&0.328 & ... &... &... &2.6 &2.1  \\ 
186235  &   1.509  &       3&0.410 & ... &... &... &1.3 &2.5  \\ 
186194  &   0.697  &       2&0.150 & -5.07 &... &9.90 &2.3 &2.1  \\ 
186268  &   0.468  &       2&0.174 & -4.86 &5. &9.61 &2.8 &2.7  \\ 
186803  &   0.689  &       9&0.254 & -4.65 &17. &9.18 &4.1 &7.2  \\ 
186853  &   0.669  &       6&0.200 & -4.81 &22. &9.52 &3.2 &4.1  \\ 
187154  &   0.624  &       8&0.288 & -4.53 &9. &8.86 &4.7 &14.3  \\ 
187101  &   0.584  &       5&0.141 & -5.12 &... &9.93 &2.2 &2.6  \\ 
hip98105  &   1.340  &       1&0.826 & ... &... &... &2.3 &2.2  \\ 
187456  &   1.058  &       6&0.361 & ... &... &... &1.6 &2.1  \\ 
188474  &   1.025  &       5&0.319 & ... &... &... &1.6 &2.1  \\ 
188559  &   1.050  &       6&0.449 & ... &... &... &1.6 &2.1  \\ 
190613  &   0.629  &       5&0.145 & -5.10 &... &9.90 &2.2 &2.4  \\ 
190647  &   0.743  &       4&0.147 & -5.10 &... &9.90 &2.1 &2.6  \\ 
190125  &   0.708  &      10&0.278 & -4.60 &16. &9.07 &2.7 &8.0  \\ 
189310  &   0.896  &       3&0.381 & -4.60 &23. &9.06 &3.1 &5.3  \\ 
191760  &   0.668  &       4&0.145 & -5.11 &... &9.91 &2.2 &2.6  \\ 
hip99764  &   1.351  &       1&1.402 & ... &... &... &3.9 &2.2  \\ 
192961  &   1.167  &       5&0.528 & ... &... &... &1.6 &2.1  \\ 
hip100356  &   1.594  &       1&1.446 & ... &... &... &5.1 &2.5  \\ 
193567  &   0.788  &       3&0.158 & -5.05 &... &9.87 &2.1 &2.1  \\ 
193995  &   0.716  &       3&0.140 & -5.15 &... &9.96 &2.0 &2.6  \\ 
195284  &   0.709  &       2&0.301 & -4.56 &14. &8.95 &2.8 &9.5  \\ 
195145  &   0.738  &       6&0.178 & -4.94 &34. &9.73 &2.2 &2.1  \\ 
196877  &   1.324  &       7&0.767 & ... &... &... &2.2 &2.1  \\ 
197210  &   0.711  &       6&0.183 & -4.90 &30. &9.68 &2.2 &2.1  \\ 
197069  &   0.579  &       5&0.168 & -4.92 &16. &9.72 &2.7 &2.6  \\ 
197499  &   0.576  &       5&0.148 & -5.05 &... &9.88 &2.4 &2.6  \\ 
197818  &   0.620  &       8&0.159 & -4.99 &23. &9.81 &2.5 &2.4  \\ 
197823  &   0.759  &       4&0.222 & -4.79 &28. &9.48 &2.4 &3.7  \\ 
hip103019  &   1.332  &       5&1.244 & ... &... &... &3.5 &2.1  \\ 
199065  &   0.658  &       4&0.246 & -4.65 &15. &9.18 &4.0 &7.8  \\ 
199704  &   0.955  &       5&0.315 & ... &... &... &2.7 &2.1  \\ 
200733  &   0.519  &       3&0.135 & -5.16 &... &9.99 &2.2 &2.9  \\ 
200869  &   0.851  &       6&0.157 & -5.07 &... &9.90 &2.1 &2.1  \\ 
201796A  &   0.654  &       7&0.313 & -4.49 &9. &8.77 &5.1 &14.7  \\ 
201796B  &   0.654  &       7&0.317 & -4.49 &9. &8.74 &5.2 &15.2  \\ 
201757  &   0.714  &       6&0.140 & -5.14 &... &9.96 &2.1 &2.6  \\ 
202041  &   0.578  &       3&0.151 & -5.03 &... &9.86 &2.4 &2.6  \\ 
202206  &   0.714  &      10&0.214 & -4.78 &25. &9.47 &2.4 &4.1  \\ 
202457  &   0.689  &       6&0.146 & -5.10 &... &9.90 &2.2 &2.1  \\ 
202746  &   1.010  &       4&0.998 & ... &... &... &1.6 &11.9  \\ 
198477  &   0.900  &       6&0.227 & -4.88 &41. &9.64 &2.4 &2.1  \\ 
203432  &   0.741  &       6&0.186 & -4.90 &33. &9.68 &2.3 &2.1  \\ 
203413  &   1.063  &       5&0.545 & ... &... &... &1.6 &2.1  \\ 
203850  &   0.924  &       4&0.243 & ... &... &... &2.4 &2.1  \\ 
204313  &   0.697  &       8&0.161 & -5.00 &... &9.82 &2.5 &2.1  \\ 
204807  &   0.717  &       2&0.197 & -4.85 &28. &9.59 &2.3 &2.1  \\ 
204941  &   0.878  &       4&0.218 & -4.88 &40. &9.65 &2.3 &2.1  \\ 
205067  &   0.656  &       4&0.158 & -5.01 &... &9.83 &2.4 &2.3  \\ 
205045  &   0.591  &       6&0.137 & -5.16 &... &9.99 &2.1 &2.6  \\ 
205158  &   0.596  &      13&0.130 & -5.24 &... &10.09 &2.0 &2.6  \\ 
205310  &   1.379  &       3&0.856 & ... &... &... &2.3 &2.2  \\ 
205739  &   0.546  &      15&0.139 & -5.12 &... &9.93 &2.2 &2.8  \\ 
205860  &   0.500  &       4&0.142 & -5.09 &... &9.91 &2.3 &3.0  \\ 
206025  &   0.555  &      14&0.134 & -5.18 &... &10.01 &2.1 &2.7  \\ 
206255  &   0.727  &       6&0.131 & -5.22 &... &10.06 &2.0 &4.3  \\ 
206683  &   0.657  &       7&0.150 & -5.07 &... &9.89 &2.3 &2.3  \\ 
207315  &   0.401  &       2&0.170 & -4.89 &3. &9.66 &2.2 &3.0  \\ 
hip107772  &   1.328  &       1&0.848 & ... &... &... &2.4 &2.1  \\ 
207970  &   0.733  &       8&0.155 & -5.05 &... &9.88 &2.1 &2.1  \\ 
207496  &   1.000  &       5&0.603 & ... &... &... &3.9 &5.9  \\ 
208573  &   1.035  &       3&0.475 & ... &... &... &1.6 &2.1  \\ 
208704  &   0.640  &       3&0.152 & -5.05 &... &9.88 &2.3 &2.4  \\ 
209566  &   0.758  &       5&0.156 & -5.05 &... &9.88 &2.1 &2.1  \\ 
209659  &   0.609  &       6&0.142 & -5.12 &... &9.93 &2.2 &2.6  \\ 
209913  &   0.763  &       6&0.167 & -4.99 &40. &9.81 &2.2 &2.1  \\ 
210193  &   0.660  &       8&0.172 & -4.93 &25. &9.73 &2.7 &2.3  \\ 
211366  &   0.659  &      10&0.172 & -4.93 &25. &9.72 &2.7 &2.3  \\ 
211369  &   0.961  &       5&0.463 & ... &... &... &3.3 &5.0  \\ 
211970  &   1.329  &       8&1.184 & ... &... &... &3.3 &2.1  \\ 
hip110655  &   1.312  &       3&1.282 & ... &... &... &3.7 &2.1  \\ 
213241  &   0.463  &       5&0.149 & -5.03 &... &9.85 &2.3 &3.0  \\ 
213717  &   0.817  &       9&0.189 & -4.93 &40. &9.72 &2.3 &2.1  \\ 
213401  &   0.669  &       2&0.151 & -5.06 &... &9.89 &2.3 &2.5  \\ 
213941  &   0.670  &       4&0.178 & -4.91 &26. &9.69 &2.8 &2.2  \\ 
214100  &   1.411  &       3&0.831 & ... &... &... &2.2 &2.3  \\ 
214385  &   0.640  &       5&0.155 & -5.03 &... &9.85 &2.4 &2.4  \\ 
214691  &   0.735  &       5&0.149 & -5.08 &... &9.91 &2.1 &2.1  \\ 
215456  &   0.636  &       7&0.133 & -5.20 &... &10.04 &2.0 &2.4  \\ 
216054  &   0.741  &       8&0.176 & -4.94 &35. &9.74 &2.2 &2.1  \\ 
216133  &   1.452  &       5&1.350 & ... &... &... &3.6 &2.4  \\ 
hip113850  &   1.643  &       2&1.326 & ... &... &... &3.6 &2.5  \\ 
218511  &   1.201  &       7&1.208 & ... &... &... &1.6 &4.3  \\ 
218572  &   1.004  &       5&0.461 & ... &... &... &1.6 &3.9  \\ 
218760  &   0.966  &       2&0.321 & ... &... &... &2.6 &2.1  \\ 
219011  &   0.727  &       6&0.149 & -5.09 &... &9.91 &2.1 &2.1  \\ 
hip114719  &   1.449  &       4&1.325 & ... &... &... &3.6 &2.4  \\ 
219533  &   0.684  &       3&0.138 & -5.16 &... &9.98 &2.1 &2.4  \\ 
219556  &   0.759  &       3&0.158 & -5.04 &... &9.86 &2.1 &2.1  \\ 
219495  &   1.130  &       2&0.863 & ... &... &... &1.6 &4.3  \\ 
219509  &   1.049  &       2&0.311 & ... &... &... &1.6 &2.1  \\ 
219764  &   1.154  &      12&1.089 & ... &... &... &1.6 &5.1  \\ 
220256  &   0.853  &       4&0.184 & -4.96 &45. &9.77 &2.2 &2.1  \\ 
220426  &   0.581  &       4&0.146 & -5.07 &... &9.90 &2.3 &2.6  \\ 
220476  &   0.682  &       5&0.367 & -4.42 &7. &8.54 &6.1 &17.9  \\ 
220689  &   0.603  &       5&0.159 & -4.99 &20. &9.81 &2.5 &2.5  \\ 
hip115752  &   1.272  &       2&0.809 & ... &... &... &1.6 &2.1  \\ 
220718  &   0.651  &       5&0.132 & -5.22 &... &10.05 &2.0 &2.6  \\ 
220829  &   0.906  &       3&0.166 & ... &... &... &2.1 &2.1  \\ 
220945  &   0.934  &       7&0.303 & ... &... &... &2.7 &2.1  \\ 
220981  &   0.740  &       5&0.142 & -5.14 &... &9.95 &2.1 &2.1  \\ 
hip115955  &   1.316  &       2&1.633 & ... &... &... &4.6 &2.1  \\ 
221257  &   0.636  &       3&0.166 & -4.96 &23. &9.76 &2.6 &2.4  \\ 
221275  &   0.799  &       2&0.236 & -4.77 &30. &9.45 &2.5 &2.1  \\ 
hip116491  &   1.370  &      10&1.490 & ... &... &... &4.0 &2.2  \\ 
221954  &   0.750  &       4&0.145 & -5.11 &... &9.92 &2.1 &2.6  \\ 
222422  &   0.731  &       4&0.237 & -4.72 &23. &9.35 &2.5 &4.9  \\ 
222743  &   0.559  &       4&0.153 & -5.02 &... &9.84 &2.5 &2.7  \\ 
222669  &   0.608  &       3&0.166 & -4.94 &20. &9.75 &2.6 &2.5  \\ 
223957  &   0.607  &       4&0.205 & -4.76 &14. &9.42 &3.3 &4.3  \\ 
224010  &   0.628  &       6&0.146 & -5.09 &... &9.91 &2.3 &2.4  \\ 
224022  &   0.572  &       9&0.147 & -5.06 &... &9.89 &2.3 &2.7  \\ 
224143  &   0.640  &       5&0.278 & -4.56 &11. &8.95 &4.6 &11.9  \\ 
224228  &   0.973  &       5&0.640 & ... &... &... &4.1 &7.6  \\ 
224376  &   0.486  &       3&0.186 & -4.80 &6. &9.50 &3.0 &7.5  \\ 
224433  &   0.748  &       5&0.167 & -4.99 &38. &9.80 &2.2 &2.1  \\ 
224464  &   0.465  &       2&0.158 & -4.96 &6. &9.77 &2.5 &3.0  \\ 
224538  &   0.581  &       4&0.153 & -5.02 &... &9.84 &2.4 &2.6  \\ 
224607  &   1.042  &       8&0.640 & ... &... &... &1.6 &4.9  \\ 
\enddata
\end{deluxetable}

\begin{figure}
\includegraphics[angle=0,width=\textwidth]{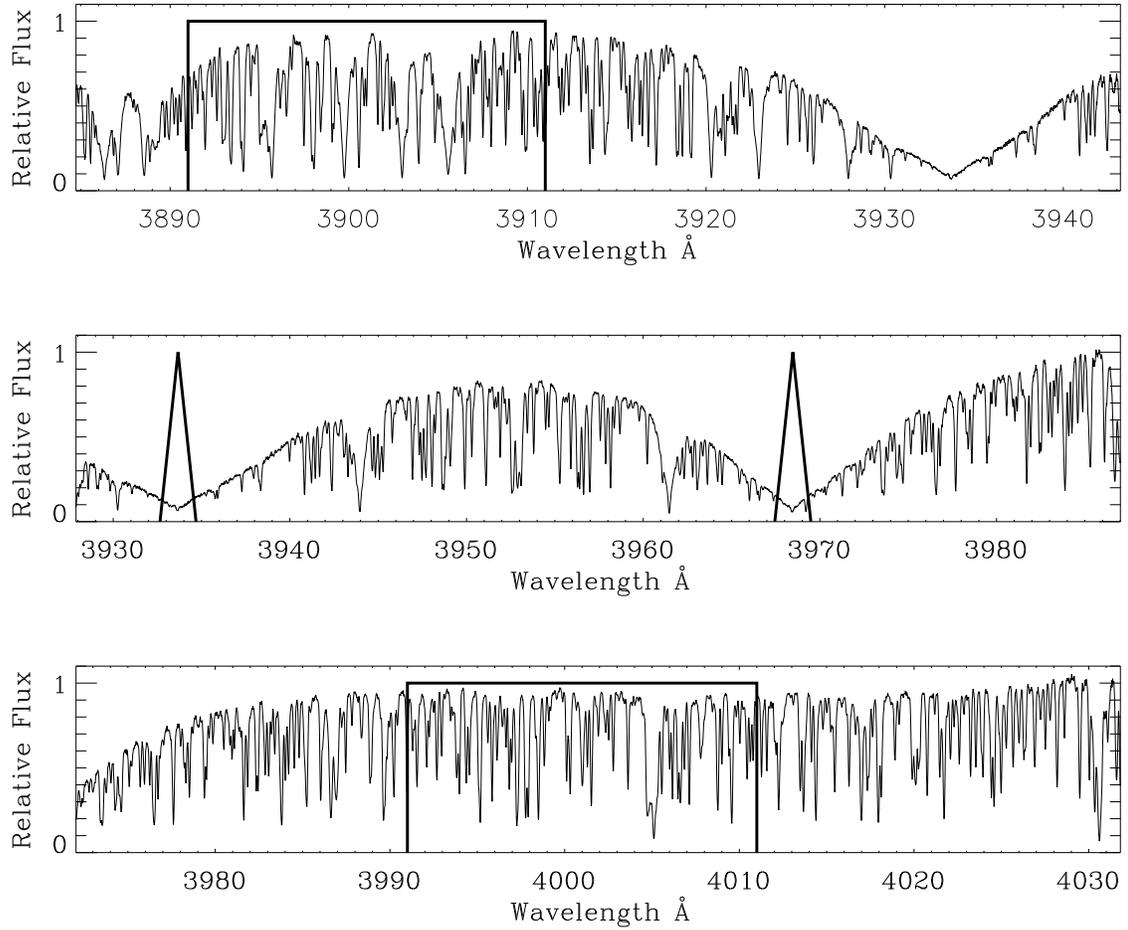}
\caption{ $V$, $K$, $H$ and $R$ channels in a representative MIKE spectrum. The relative flux is in arbitrary units. Wavelengths have been shifted to zero velocity in order to make the measurements. }
\label{fig1}
\end{figure}

\begin{figure}
\includegraphics[angle=90,width=\textwidth]{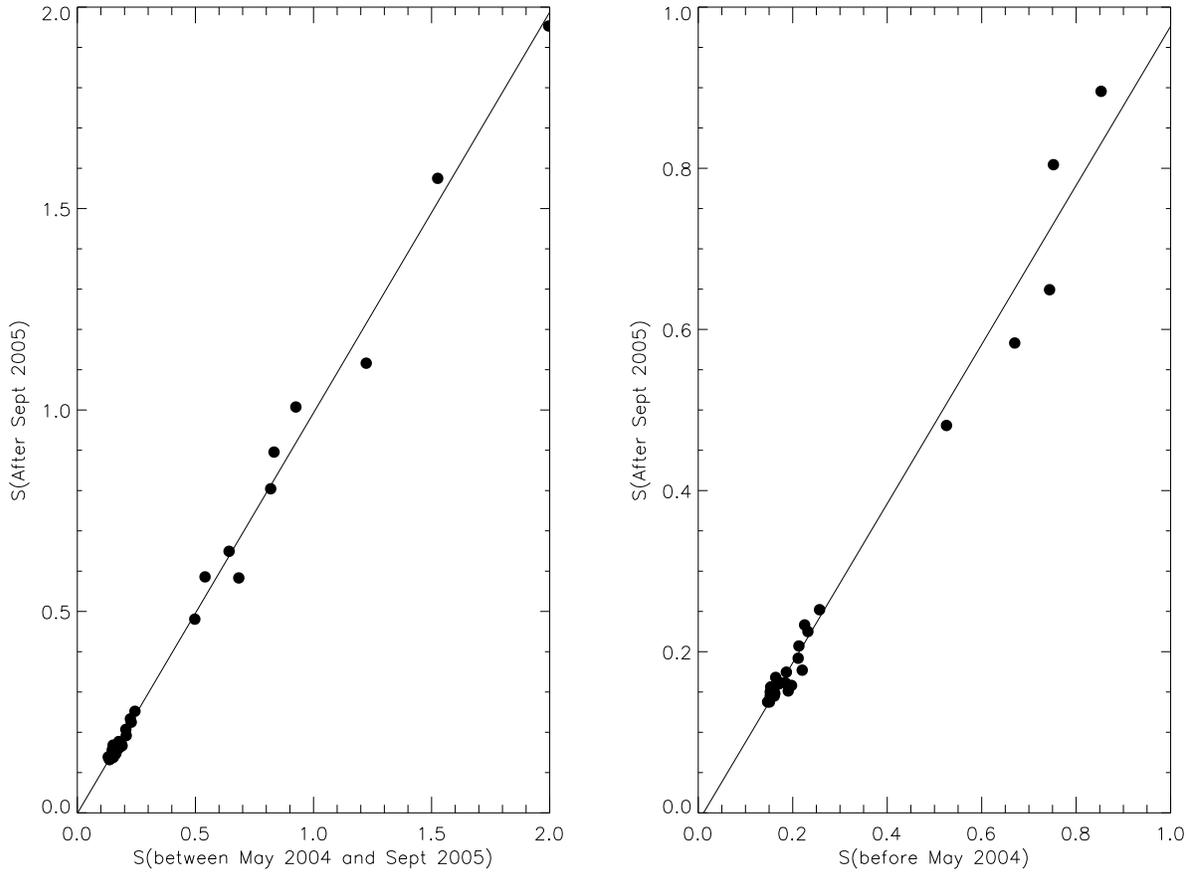}
\caption{ Comparison between \smi measurements taken in different epochs. In the left panel, filled circles correspond to measured S values using data taken between May 2004 and Sept. 2005 vs. S measured using data taken after Sept. 2005. The solid line corresponds to a linear regression, yielding a slope of 0.995 and an intercept of -0.002. In the right panel,  filled circles correspond to measured S values using data taken before May 2004 vs. measured using data taken after Sept. 2005. The solid line corresponds to a linear regression, yielding a slope of 0.987 and an intercept of  -0.011.}
\label{fig2}
\end{figure}

\begin{figure}
\includegraphics[angle=0,width=\textwidth]{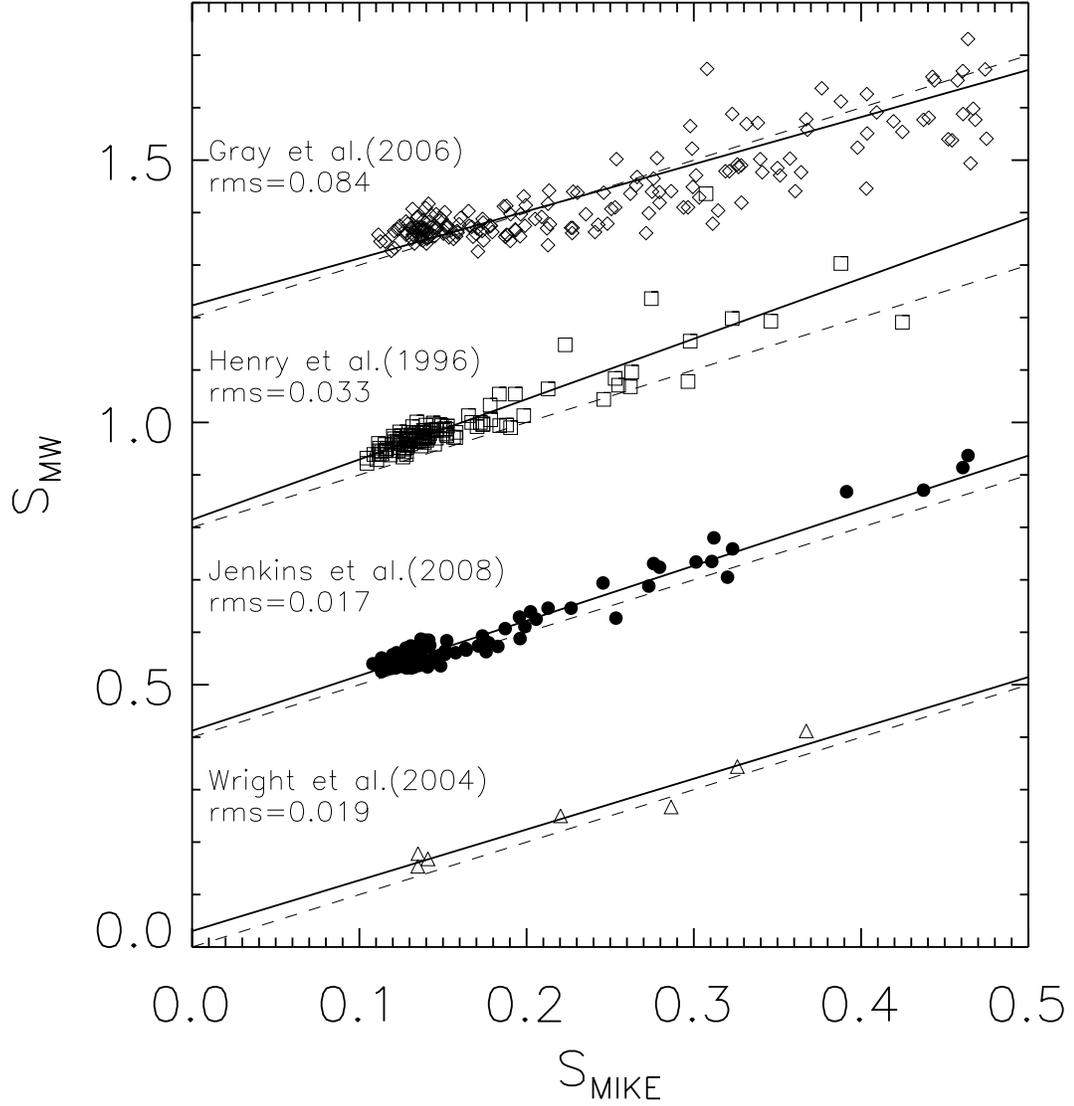}
\caption{ Comparison between \smi measurements with previous measurements that are already calibrated to the \smw system. Solid lines correspond to linear regressions for each of the datasets, while the dashed lines show relations of slope unity. For clarity, the Jenkins et al (2008), Henry et al. (1996) and Gray et al. (2006) datasets have an offset of 0.4, 0.8 and 1.2 respectively}
\label{fig3}
\end{figure}

\begin{figure}
\includegraphics[angle=0,width=\textwidth]{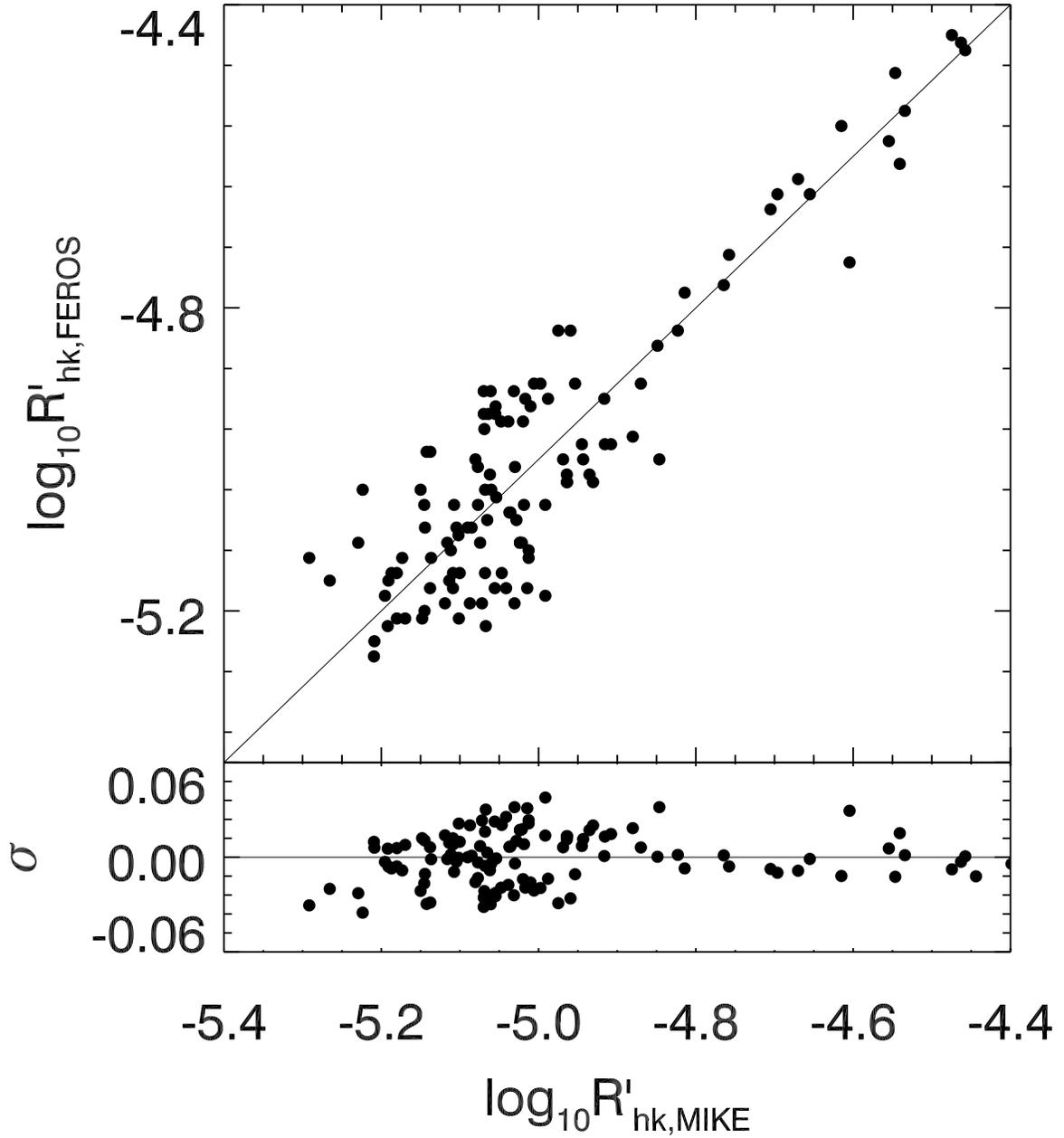}
\caption{ Upper panel : Comparison between \logrhk indices obtained from MIKE spectra in this work and those obtained from FEROS spectra by Jenkins et al. (2008). The line denotes a 1:1 relationship. Lower panel : the level of difference between both samples, or $\sigma$ values vs. our \logrhk indicies. The line denotes 0 level of difference between the samples.}
\label{fig4}
\end{figure}

\begin{figure}
\includegraphics[angle=90,width=\textwidth]{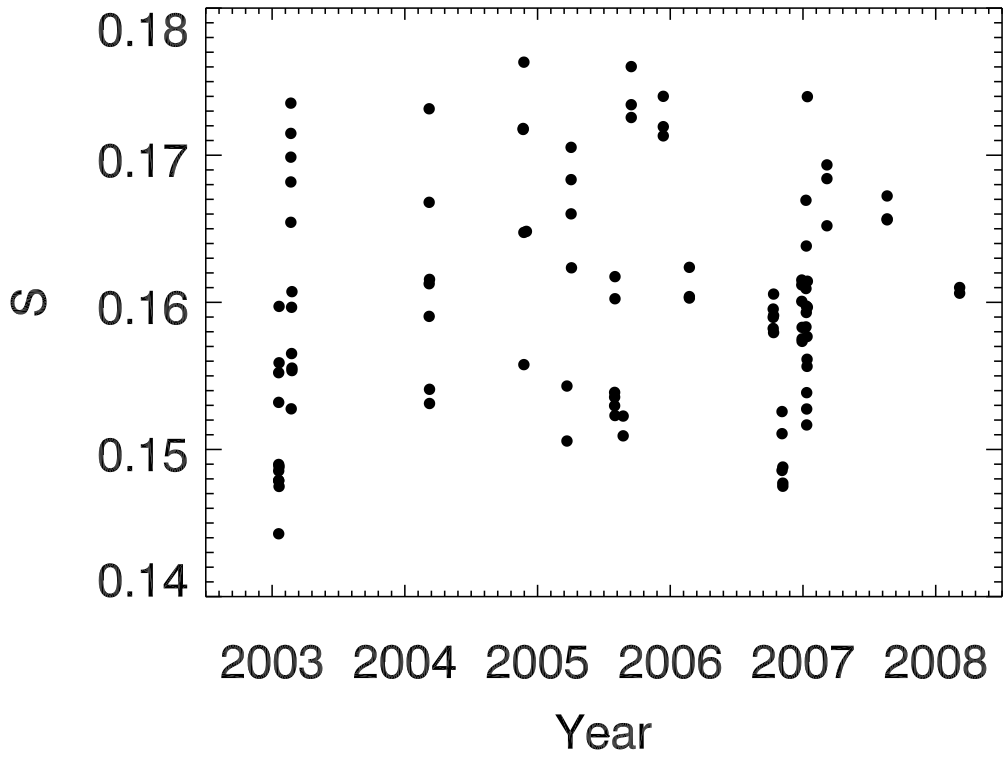}
\caption{ $\tau$ Ceti S-values from Magellan/MIKE. Observations have a  standard deviation of 4.9\%}
\label{fig5}
\end{figure}

\begin{figure}
\includegraphics[angle=90,width=\textwidth]{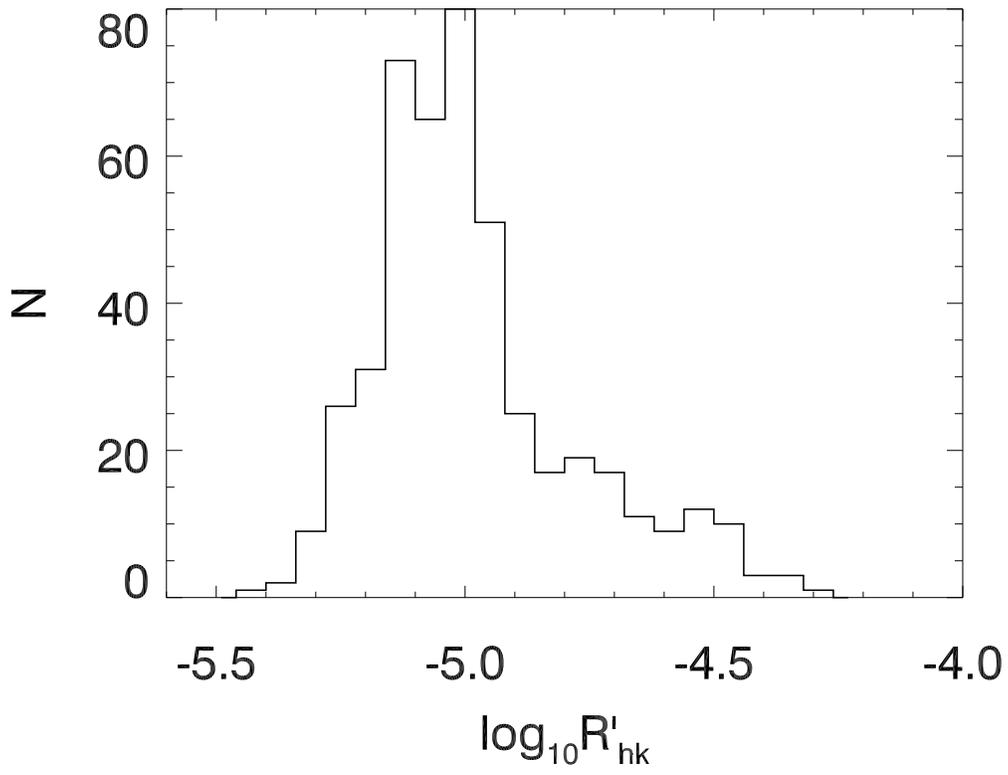}
\caption{ Distribution of the chromospheric activity parameter \logrhk for the target stars of the Magellan Planet Search Program. The bulk of stars are inactive, with a peak at values below -4.9. }
\label{fig6}
\end{figure}

\end{document}